\documentclass[11pt, twoside]{article}
\usepackage{amsfonts}
\usepackage{amsmath}
\usepackage{amssymb}
\usepackage{amsthm}
\usepackage{latexsym}
\usepackage{natbib}
\usepackage{graphicx}
\usepackage{xcolor}
\usepackage{fancyhdr}
\usepackage{fancybox}
\usepackage[T1]{fontenc}
\usepackage[scaled]{helvet}

\title{CircSiZer: an exploratory tool for circular data}
\author{M. Oliveira*, R. M. Crujeiras and A. Rodr\'iguez--Casal\\ \small{Department of Statistics and Operations Research}\\ \small{University of Santiago de Compostela (Spain)}}
\date{}
\begin{document}
\maketitle

\begin{center}
\textbf{Abstract}
\end{center}
\begin{quotation}
Smoothing methods and SiZer (SIgnificant ZERo crossing of the derivatives) are useful tools for exploring significant underlying structures in data samples.  An extension of SiZer to circular data, namely CircSiZer, is introduced. Based on scale--space ideas, CircSiZer presents a graphical device to assess which observed features are statistically significant, both for density and regression analysis with circular data. The method is intended for analyzing the behavior of wind direction in the atlantic coast of Galicia (NW Spain) and how it has an influence over wind speed. The performance of CircSiZer is also checked with some simulated examples.\\

\end{quotation}

\vspace*{5cm}
\begin{quotation}
\noindent
\textbf{*Corresponding author:}\\
Mar\'ia Oliveira\\
Department of Statistics and Operations Research\\
Faculty of Mathematics - University of Santiago de Compostela (Spain)\\
e--mail: maria.oliveira@usc.es
\end{quotation}

\section{Introduction}

\begin{figure}[h!]
\begin{center}
\includegraphics[width=6.5cm]{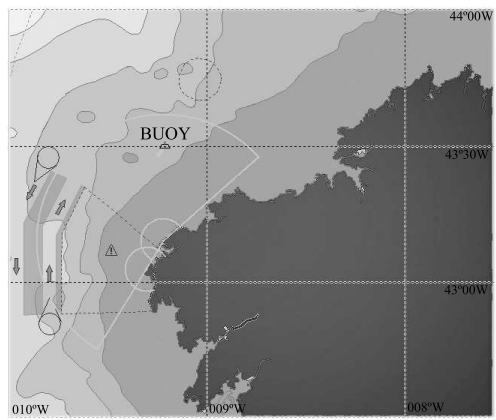}
\caption{Atlantic coast of Galicia (NW Spain). The plot shows the marine traffic control area (arrows indicate the directions that ships must follow), whithin the influence area of two major lighthouses (white lines). The buoy registering the data is located NE from the traffic control area at longitude -0.210E and latitude 43.500N.}
\label{coast_map}
\end{center}
\end{figure}

Coastal and marine ecosystems suffer from a variety of threats due to human and industrial activity, being these ecosystems specially vulnerable to oil spills and toxic dumping. Specifically, the atlantic coast of Galicia (NW Spain) has suffered two major ship accidents which caused serious environmental and ecological damages: the burning of a cargo ship named Cas\'on in 1987, and the oil spill of the Prestige tanker, in 2002. In the first accident, the strong winds caused a displacement of the cargo, and the corrosive and toxic chemical flamable products transported by Cas\'on exploded and burned, while the ship was handling on a dock. Also because of the highly variable and strong winds in the area during a storm, the Prestige oil tanker sank in front of the Galician coast causing the largest environmental disaster in the atlantic coast of the Iberian peninsula with the spill of more than seventy thousand tonnes of fuel. Despite the occurence of these serious accidents, this area is still on the course of most cargo vessels and tankers navigating from the north of Europe to the Mediterranean Sea, Africa or America. As it is shown in Figure \ref{coast_map}, there exists a marine traffic control zone, which regulates the sailing direction and distance from the coast. A buoy anchored in the area (see Figure \ref{coast_map}) provides hourly collected wind speed and wind direction, being the measurements of this latter variable a set of circular data.

Circular data are data that can be represented as directions in a unit circle (see Jammalamadaka and SenGupta 2001, for an extensive review on this topic). This type of data arise quite frequently in many natural and physical sciences, such as, in marine sciences, where the study of ocean currents and winds is extremely important for marine operations involving navigation, search and rescue at sea or pollutants dispersion in the ocean. 

\begin{figure}[h!]
\begin{center}
\includegraphics[width=5cm]{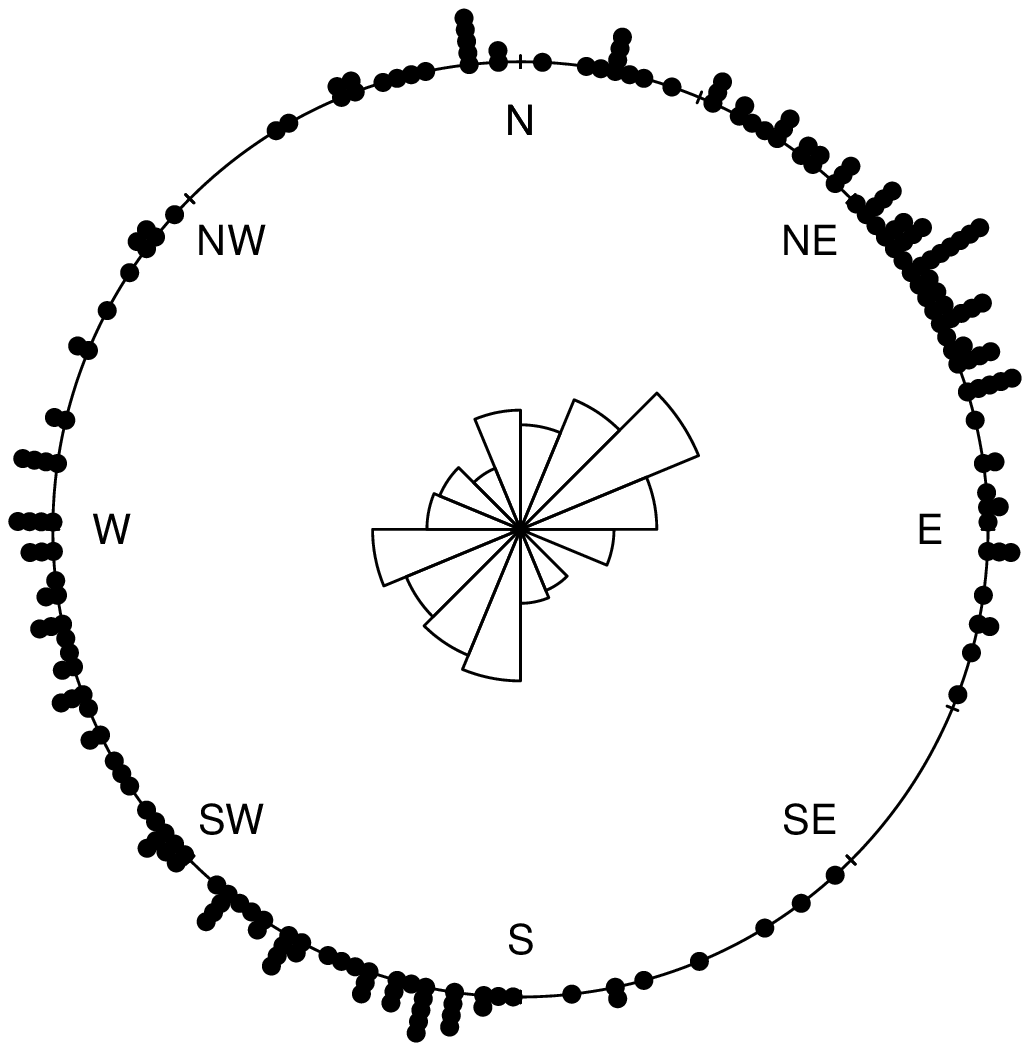}
\includegraphics[width=5cm]{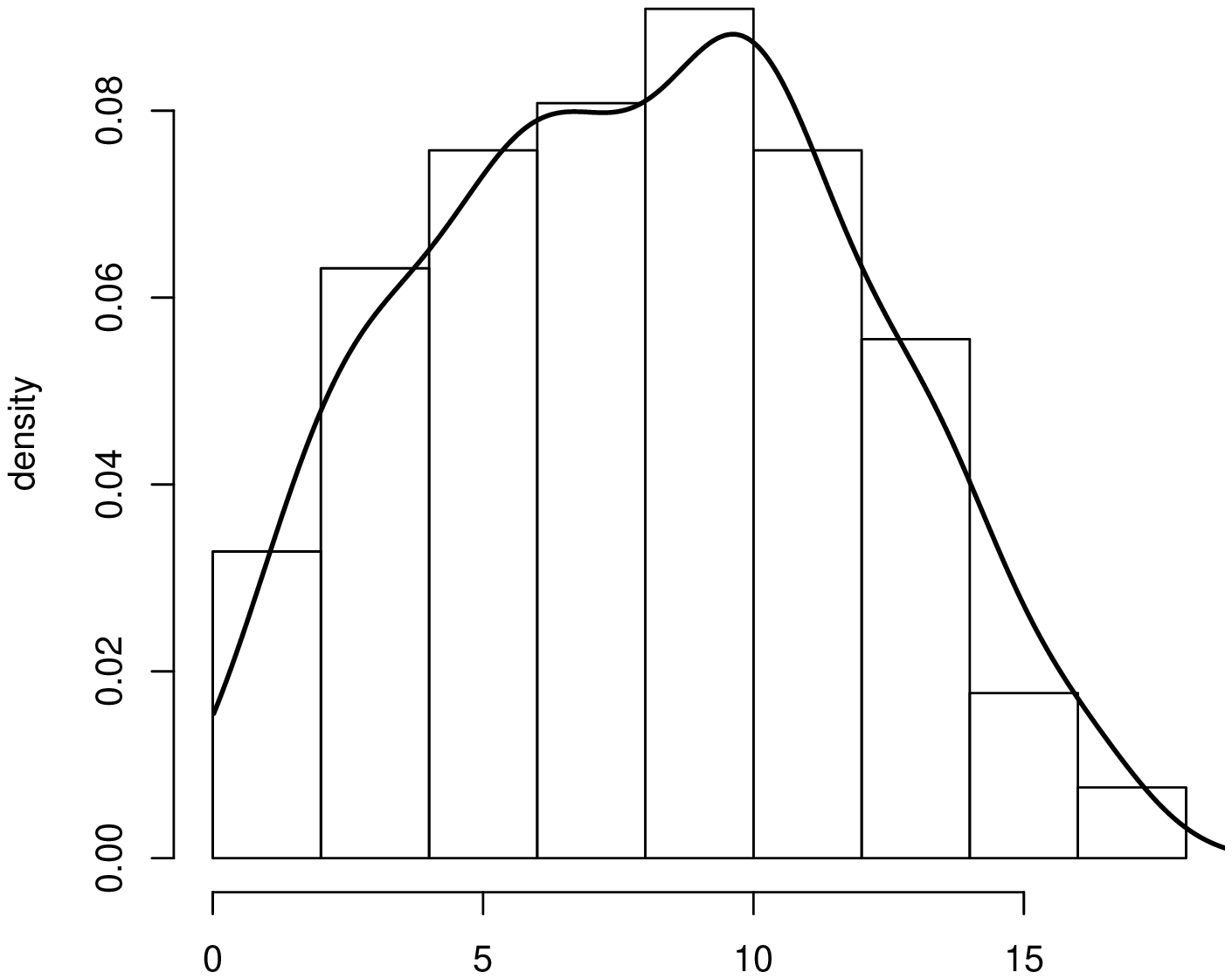}
\caption{Descriptive plots for wind direction (left, rose diagram) and wind speed (right, histograma and kernel density estimator). Wind direction is measured in angles and represented in the circumference in clockwise sense, starting from N direction. Wind speed is measured in m/s.}
\label{data_plot}
\end{center}
\end{figure}

Within this context, the goal of this work is twofold: firstly, from a practical point of view, the main aim is to describe the wind pattern in the Galician coast during winter season, focusing on the most significative wind directions and their relation with wind speed. For that purpose, a meteorological data set consisting of wind direction and wind speed measurements will be considered (see Figure \ref{data_plot} for descriptive plots). Secondly, in order to achieve this previous goal, the CircSiZer, a new exploratory tool based on nonparametric kernel density and regression estimators for circular data will be introduced.

Density estimation from a sample of circular data, as well as regression estimation when the explanatory variable is circular, is indeed an interesting statistical problem in a variety of applied fields. From a nonparametric perspective, density and regression estimation can be approached by using local smoothers based on kernel functions. Kernel density estimation for the general case of spherical data was studied by Hall et. al (1987) and nonparametric kernel methods for regression estimation for a circular explanatory variable and a linear response have been recently introduced by  Di Marzio et al. (2009). As in any nonparametric procedure, kernel methods depend on a smoothing parameter or bandwidth, which can be data--driven selected or chosen by the researcher (see Oliveira et al. 2012a, for a comparison on the existing bandwidth selectors for density estimation). The bandwidth controls the global aspect of the estimator and its dependence on the sample. Given that an unsuitable smoothing parameter may provide a misleading estimate of the density or regression curve, the assessment of the statistical significance of observed features through the smoothed curve should be required for not compromising the extracted conclusions.

The SiZer method, developed by Chaudhuri and Marron (1999) for linear data, provides a means of circumventing the smoothing parameter selection and, at the same time, allows for the assessment of statistically significant features in the data structure. The original SiZer is a visualization method based on nonparametric curve estimates. SiZer addresses the question of which features observed in a smoothed curve are really present, or represent an important underlying structure, and not simply artifacts of the sampling noise from a scale--space perspective. In the nonparametric curve estimation context, the scale--space framework is given by a family of kernel smoothers indexed by the bandwidth parameter. SiZer considers a wide range of bandwidths, which avoids the problem of bandwidth selection, whilst peaks and troughs are identified by finding the regions of significant gradient (zero crossings of the derivative), presenting this information in a simple visual way by the SiZer map.

Several adaptations of SiZer have been proposed in the statistical literature, making it possible to extend this graphical tool to a variety of contexts such as local likelihood (Li and Marron, 2005), dependent data (Rondonotti et al. 2007) and survival data (Marron and de U\~na \'Alvarez, 2004), among others. SiZer for linear variables has been successfully applied in many different scientific fields. For example, Rudge (2008) uses this method to find peaks in geochemical distributions; Sonderegger et al. (2009) consider SiZer to detect threshold in ecological data and Ryd\'en (2010) applies SiZer to determine a possible increasing trend in hurricane activity in the North Atlantic.

In the special setting of circular data, both for kernel density and regression estimation, the adaptation of SiZer ideas must take into account the nature of the data. This particular scenario involves, specifically: (1) the assessment of the variability in the derivatives of circular kernel estimators, both for density and regression, through the computation of standard deviations and appropriate quantiles; (2) the development of a suitable visualization device to facilitate the practitioner the output interpretation. Bearing these premises in mind, the SiZer ideas can be fitted to the circular data setting yielding the CircSiZer plot presented in this work. The CircSiZer plot is produced using self-programmed code developed in the free software environment R (R Development Core Team, 2012).

This paper is organized as follows. Section \ref{nonparametric_estimation} provides a brief overview on kernel density estimation for circular data, and regression estimation for a circular explanatory variable and a linear response. Section \ref{CircSiZer} is devoted to the introduction of the CircSiZer plot, detailing its construction and interpretation. The performance of the new CircSiZer is illustrated with some simulated examples and real data in Section \ref{applications}. CircSiZer is used for describing the wind direction and the relation between wind speed and wind direction in the Galician coast during winter season. A brief discussion on the proposal and some final comments are provided in Section \ref{discussion}.

\section{Nonparametric curve estimation for circular data}
\label{nonparametric_estimation}
CircSiZer will be based on nonparametric estimates of the target curve, density or regression. For this purpose, this section provides a brief background on circular kernel density estimation and local linear regression, for circular explanatory variable and linear response. See Oliveira et al. (2012b) for a comprehensive review on nonparametric methods for circular data.

\subsection{Nonparametric circular density estimation}

Given a random sample of angles $\Theta_1,\Theta_2,\ldots,\Theta_n \in [0,2\pi)$ from some unknown density $f$, the kernel circular density estimator of $f$, at an angle $\theta$, is defined as:

\begin{equation}
\hat f(\theta;\nu)=\frac{1}{n} \sum_{i=1}^{n} K_{\nu}(\theta-\Theta_i), \quad 0\leq \theta < 2\pi,
\label{density_estimator}
\end{equation}
where $K_{\nu}$ is a circular kernel function with concentration parameter $\nu>0$ (see Di Marzio et al., 2009). As a circular kernel, the von Mises density can be considered. Also known as the circular Normal, the von Mises model, $vM(\mu,\kappa)$, is a symmetric unimodal distribution characterized by a mean direction $\mu\in[0,2\pi)$, and a concentration parameter $\kappa\geq 0$, with probability density function
\[
g(\theta;\mu,\kappa)=\frac{1}{2\pi I_0(\kappa)}\exp\left\{{\kappa \cos(\theta-\mu)}\right\}, \quad 0\leq \theta < 2\pi,
\]
where $I_0$ denotes the modified Bessel function of order $0$, which is just a normalizing constant ensuring the unit integral of the density. With this specific kernel, the density estimator (\ref{density_estimator}) is given by:
\[
\hat{f}(\theta;\nu)=  \frac{1}{n (2\pi) I_0(\nu)} \sum_{i=1}^{n} \exp{\left\{\nu \cos(\theta - \Theta_i)\right\}}, \quad 0\leq \theta < 2\pi,
\]
which is a mixture of von Mises distributions centered in $\Theta_i$ and with concentration parameter $\nu$.

\begin{figure}[h!]
\begin{center}
\includegraphics[width=5cm]{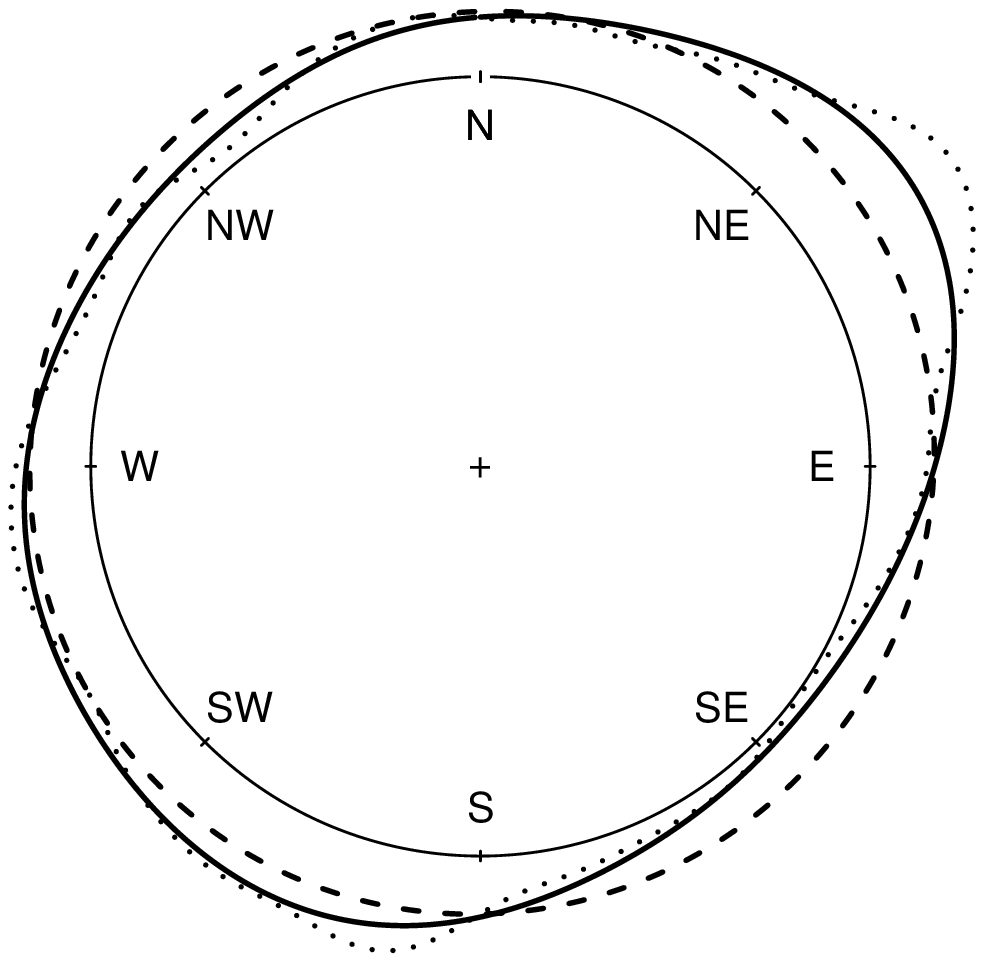}
\hspace{1cm}
\includegraphics[width=5cm]{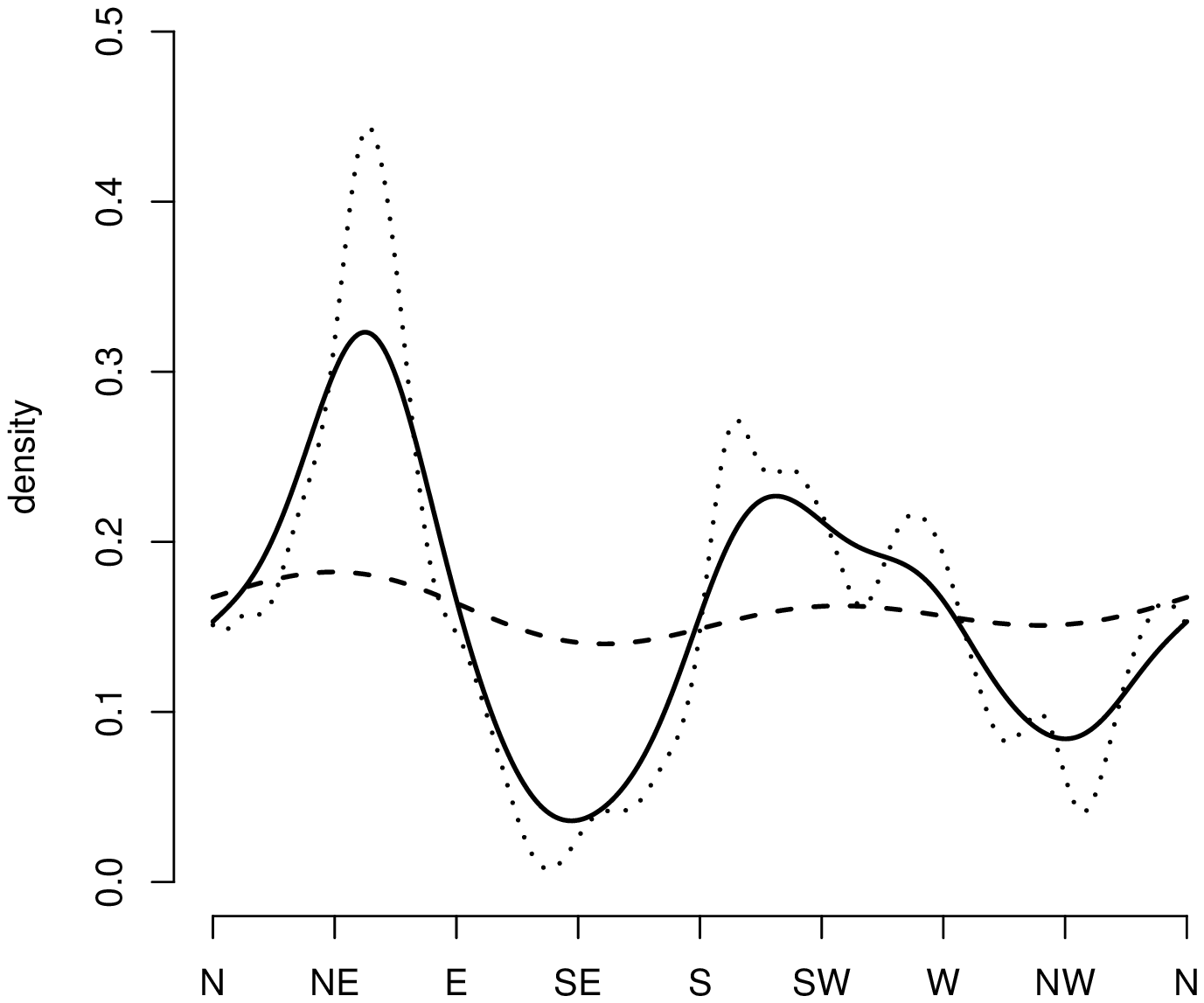}
\caption{Example of features revealed by smoothing in density estimation. Kernel circular density estimates with three different bandwidths: $\nu=1$ (dashed line), $\nu=10$ (solid line) and $\nu=60$ (dotted line) for wind direction data in circular (left panel) and linear (right panel) representations. Wind direction is represented over the circumference in clockwise sense, starting from N.}
\label{efecto_nu_datos_dens}
\end{center}
\end{figure}

A critical issue when using this estimator in practice is the choice of the smoothing parameter $\nu$. Large values of $\nu$ lead to highly variable (undersmoothed) estimators, whereas small values of $\nu$ imply low concentration of the kernel around the observations, providing oversmoothed estimators for the circular density. The effect of the smoothing parameter $\nu$ is illustrated in Figure \ref{efecto_nu_datos_dens}, where kernel density estimates for the wind direction distribution are shown. When a midrange bandwidth is used, $\nu=10$ (solid line), the estimate shows two modes suggesting that the wind comes mainly from NE and SW. However, these modes may disappear if the selected value of the bandwidth is smaller ($\nu=1$, dashed line). Also, many more modes, which are likely to be spurious sampling artifacts, appear for larger bandwidths ($\nu=60$, dotted line). A crucial issue is then how to choose the bandwidth. There are several approaches to the problem of chosing the smoothing parameter in this setting (see, e.g., Hall et al. 1987 and Oliveira et al. 2012a). Usually, the bandwidth parameter is selected in order to minimize some error criterion, such as the mean integrated squared error between the density estimate and the unknown true density. Although based on nonparametric kernel circular density (and regression) estimation, the goal of this paper is to identify which observed features are ``really there'', avoiding the selection of an ``optimal'' bandwidth parameter.

\subsection{Nonparametric circular--linear regression estimation}

\begin{figure}
\begin{center}
\includegraphics[width=5cm]{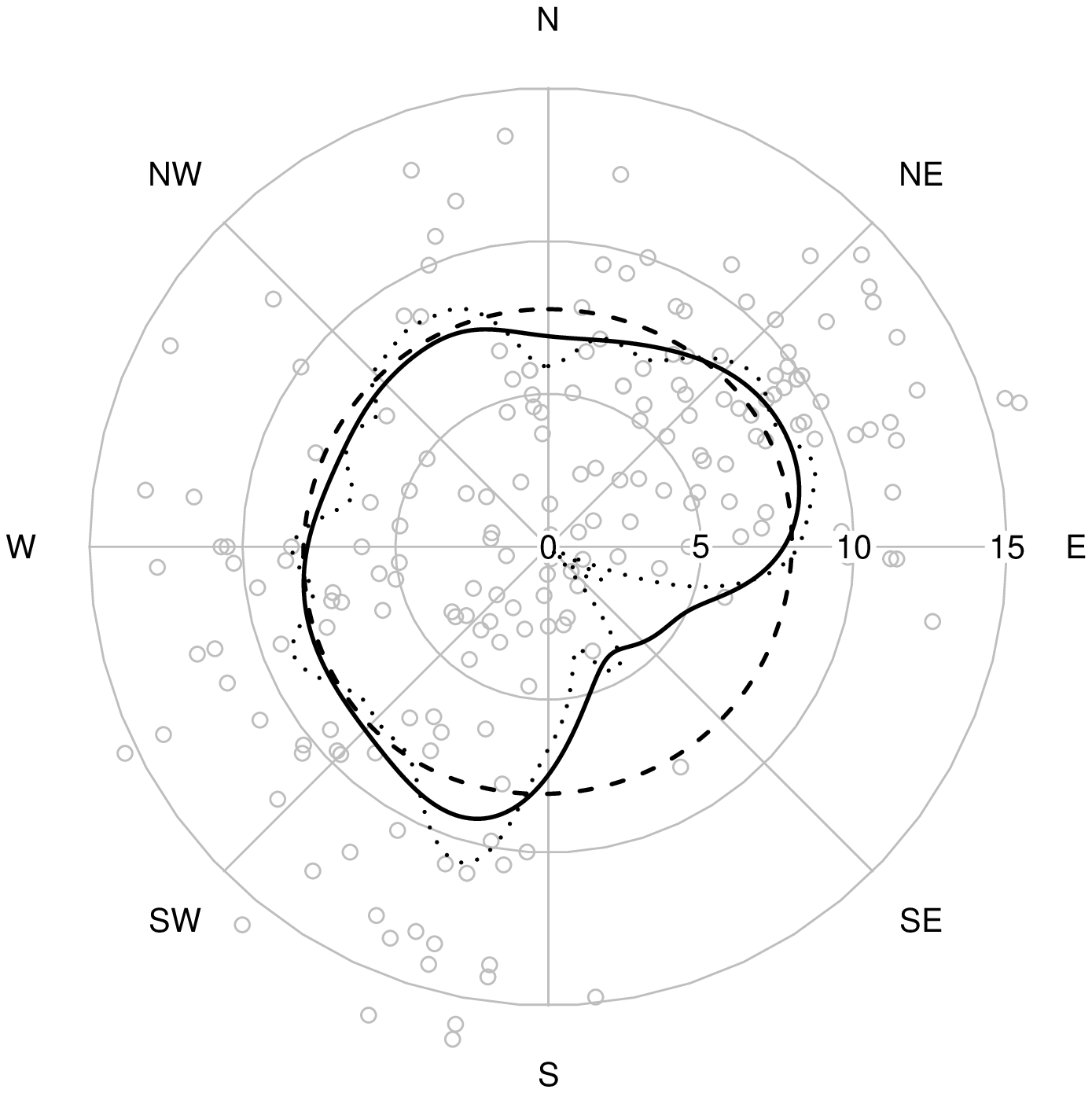}
\hspace{1cm}
\includegraphics[width=5cm]{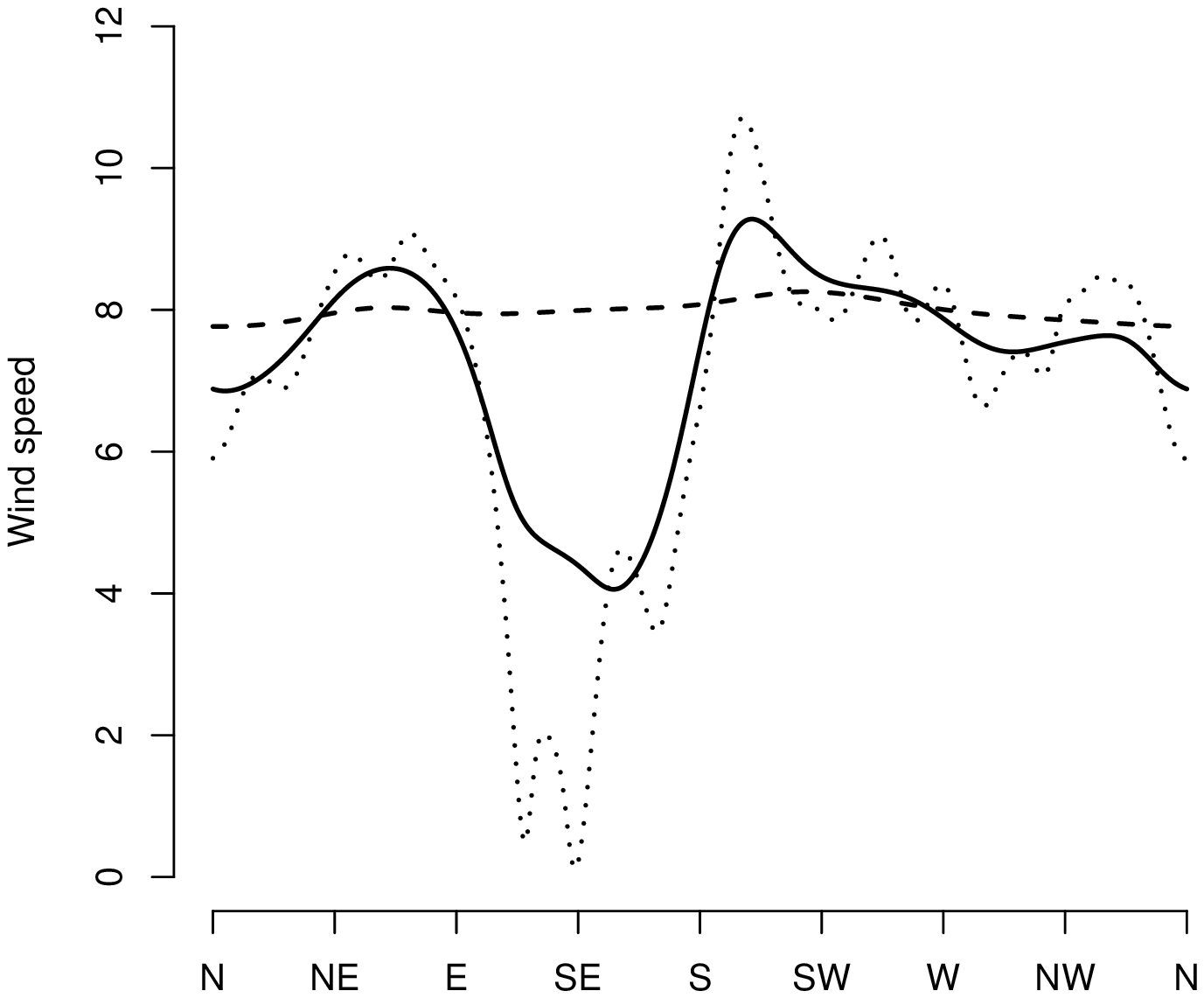}
\caption{Example of features revealed by smoothing in regression estimation. Circular-linear regression estimates, with three different bandwidths: $\nu=1$ (dashed line), $\nu=10$ (solid line) and $\nu=60$ (dotted line) for wind speed and wind direction data in circular (left panel) and linear (right panel) representations. In the left plot, wind direction is represented over the circumference in clockwise sense, starting from N and wind speed is represented along the radius.}
\label{efecto_nu_datos_regre}
\end{center}
\end{figure}

Let $\left\{(\Theta_i,Y_i),\; i=1,\ldots,n\right\}$ be a random sample from $(\Theta, Y)$ a circular and a linear random variables, respectively. The relation between these variables can be modeled by
\begin{equation}
Y_i = f(\Theta_i)+\varepsilon_i, \ i=1,\ldots,n,
\label{regression_model}
\end{equation}
where, $f$ denotes now the regression function and $\varepsilon_i$ are real--valued random variables with zero mean and variance $\sigma^2$. The local circular--linear regression estimate for $f(\theta)$ and $f^{\prime}(\theta)$ at an angle $\theta$ are given by $\hat{f}(\theta;\nu)=\hat{a}$ and $\hat{f}^{\prime}(\theta;\nu)=\hat{b}$, where
\begin{equation}
(\hat{a},\hat{b})=\arg\min_{(a,b)}\sum_{i=1}^{n}K_{\nu}(\theta-\Theta_i)\left[Y_i-(a+b\sin(\theta-\Theta_i))\right]^2
\label{regression}
\end{equation}
(see Di Marzio et al., 2009 for details). In equation (\ref{regression}), $\nu$ is the smoothing parameter and $K_{\nu}$ is a circular kernel function, and as for density estimation, a von Mises kernel with concentration parameter $\nu$ is used throughout this work. With respect to the smoothing parameter, large values of $\nu$ lead to undersmoothed estimations of the regression curve, exaggerating the local features in the sample and tending to an interpolation of the data. On the other hand, small values of $\nu$ result in a global averaging, oversmoothing the local characteristics in the data. This effect can be checked on the real data example, as shown Figure \ref{efecto_nu_datos_regre}, when plotting the estimator of the regression function for the wind speed (response) taking the wind direction as a covariate. A small value of the smoothing parameter ($\nu=1$, dashed line) provides an oversmoothed estimation indicating that there is no effect of the wind direction over the wind speed. However, for an intermediate value of $\nu$ ($\nu=10$, solid line) wind speed is higher when wind comes from NE and S and lower when coming from SE. These features are also shown by the larger bandwidth ($\nu=60$, dotted line) but, in this case, the estimator seems to be substantially undersmoothed.

A simple and widely used procedure for bandwidth selection in the regression setting is cross-validation, which pursues an ``optimal'' choice of the smoothing level (see Oliveira et al. 2012b). As already commented for the kernel circular density estimation problem, in the next section, a method for exploring the different features that occur on a range of smoothing parameter values is proposed, avoiding the problem of selecting a specific smoothing parameter.

\section{CircSiZer: SiZer map for circular data}
\label{CircSiZer}

As noticed in the previous section, bandwidth selection is a critical issue for nonparametric density and regression estimations. Apart from the lack of a uniformly superior rule for that purpose, from a practical point of view, the exploration of the estimators at different smoothing degrees (for a range of reasonable bandwidth values, between oversmoothing and undersmoothing levels) will provide more in--depth information about the available data. However, significant features in the underlying data structure should be effectively disentangled from sampling artifacts. Features like peaks and valleys of a smooth curve can be characterized in terms of zero crossings of derivatives. Hence, the significance of such features can be judged from statistical significance of zero crossings or equivalently the sign changes of derivatives. This idea has been sucessfully exploited by Chaudhuri and Marron (1999) in developing a simple yet effective tool called SiZer for exploring significant structures in density and regression curves.

\begin{figure}[h!]
\begin{center}
\includegraphics[width=3.5cm]{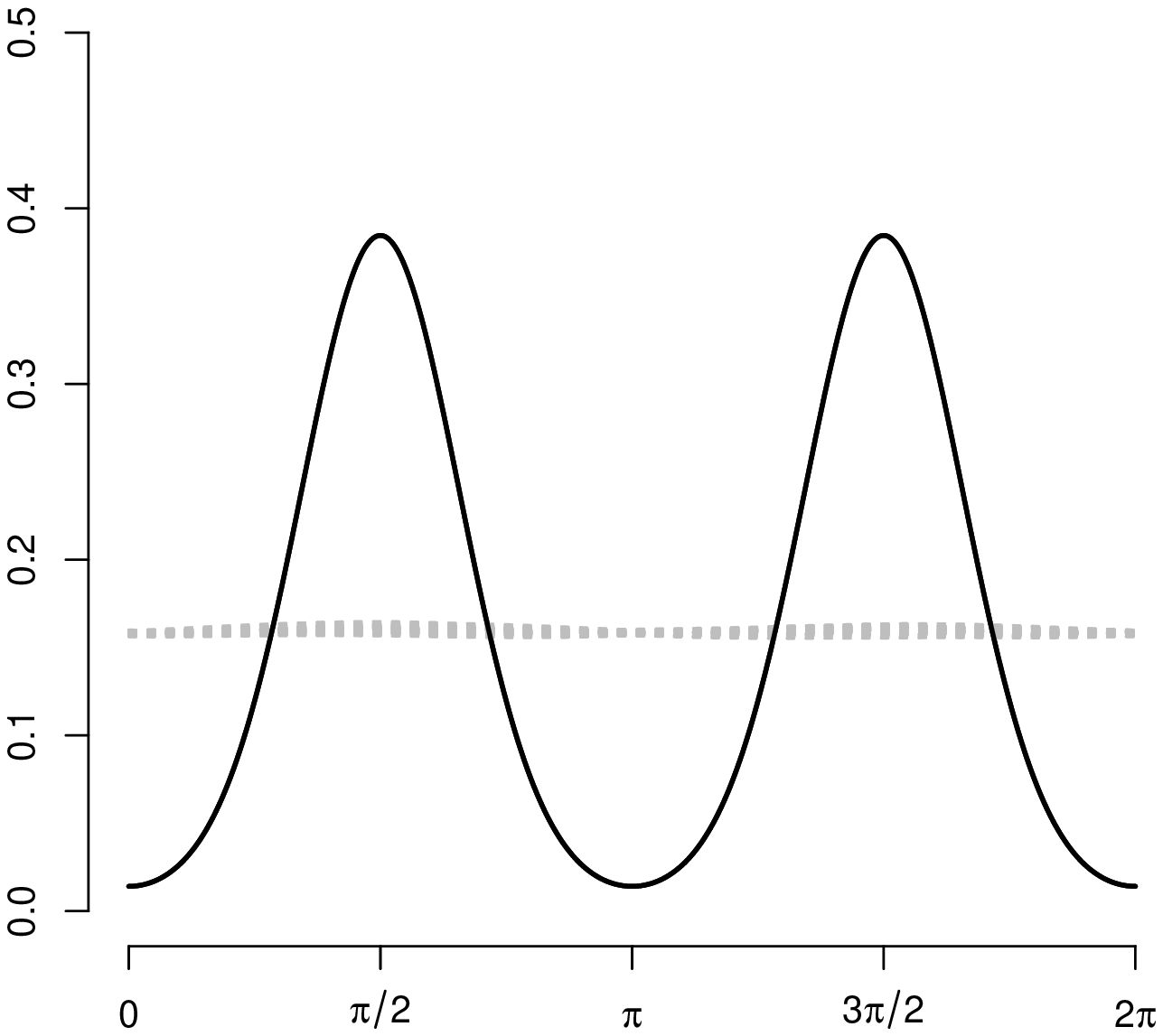}
\includegraphics[width=3.5cm]{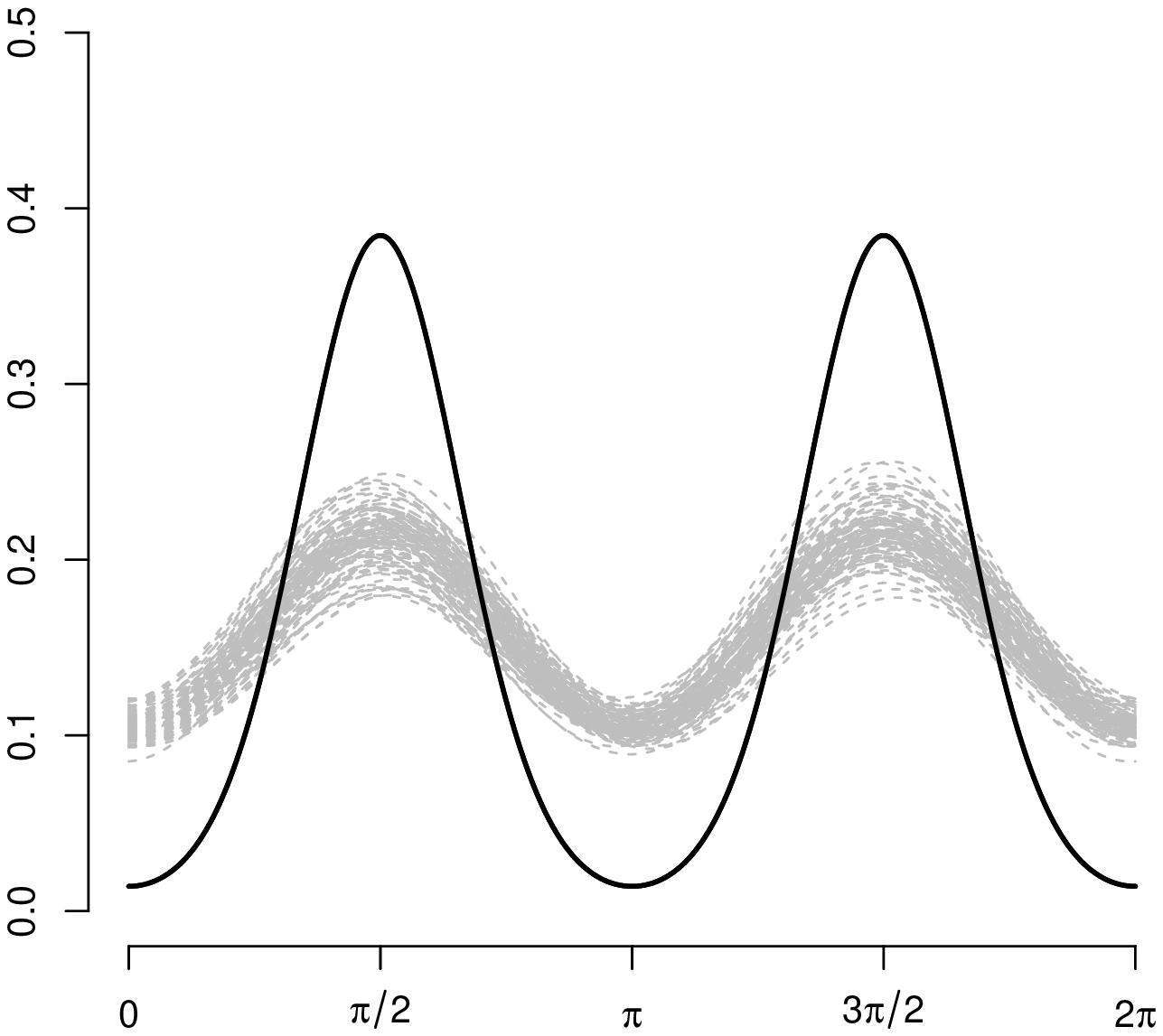}
\includegraphics[width=3.5cm]{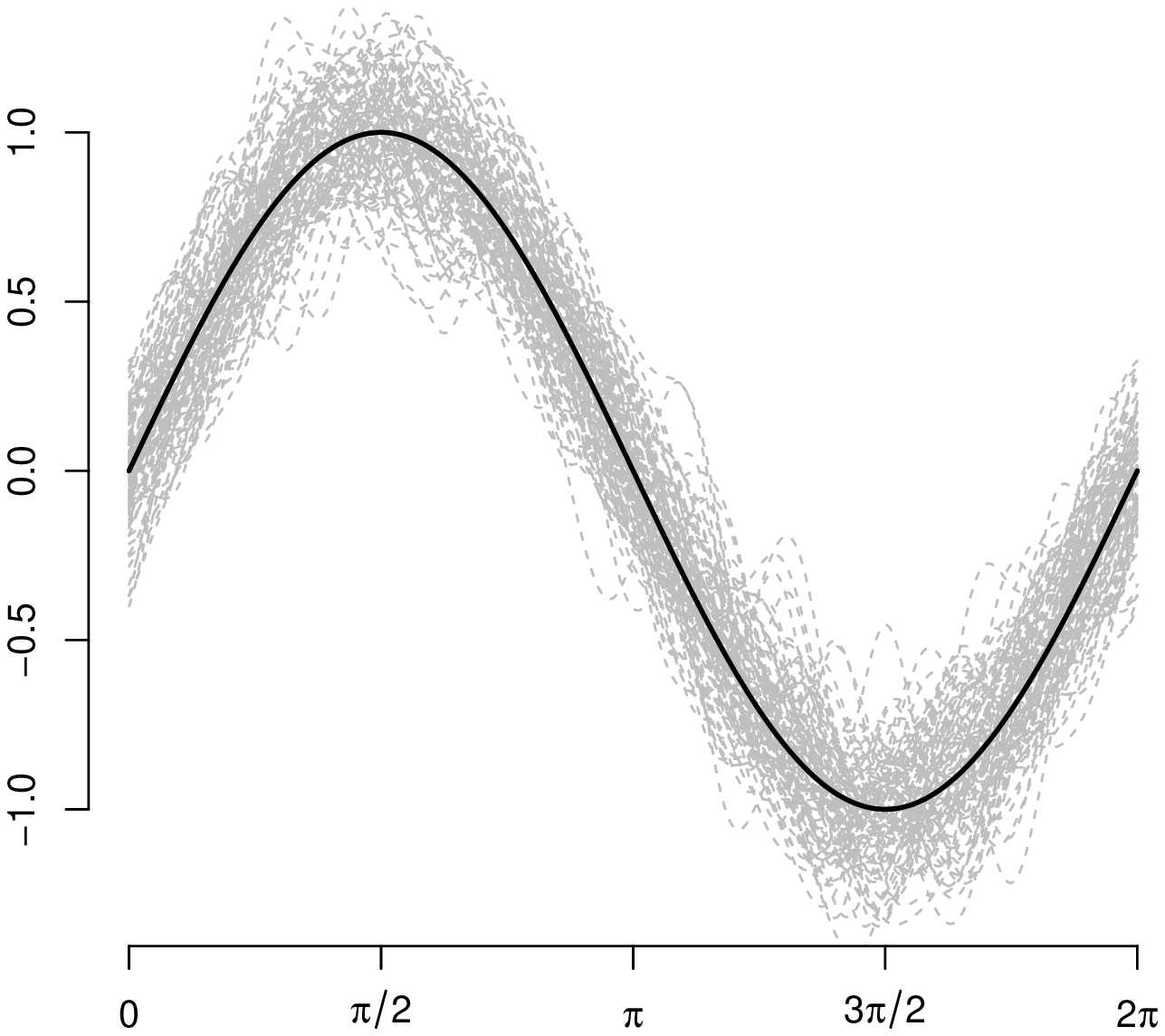}
\caption{Left and center: True density: Mixture of two von Mises in the same proportion $vM(\pi/2,4)$ and $vM(3\pi/2,4)$ (solid line), nonparametric density estimators for the mixture from 100 random sample of size 250 (gray curves) with $\nu=0.2$ and $\nu=2$, respectively. Right: Nonparametric regression estimators (gray curves) for the sine model (solid line) from 100 random samples of size 250 with $\nu=50$.}
\label{bias_problem}
\end{center}
\end{figure}

In the usual inferential approach in the statistical literature, the spotlight is placed on the true underlying curve $f$ (the regression or the density function) and doing inference on it, in particular, based on confidence bands. A crucial problem in nonparametric estimation is that $f(\theta;\nu)=\mathbb{E}(\hat{f}(\theta;\nu))$ is not necessarily equal to $f(\theta)$, involving an inherent bias specially for small values of $\nu$ (see Figure \ref{bias_problem}, left). The bias can be reduced by taking large values of $\nu$, but in this case the estimator is highly variable, depending strongly on the data sample (see Figure \ref{bias_problem} right). Chaudhuri and Marron (1999) avoid the bias-variance trade off problem by adopting the scale--space ideas which naturally lead to making inference on the smoothed curve $f(\cdot;\nu)$ rather than on the curve $f$. It should be noted that, for small values of $\nu$, the smoothed curve $f(\cdot;\nu)$ can be very different from $f$. However if $\nu$ is within a reasonable range, $f(\cdot;\nu)$, which can be thought as the curve at a resolution level $\nu$, shows the same valley-peaks structure as $f$ (see Figure \ref{bias_problem}, center).

Thus, in order to assess the significance of features such as peaks and valleys, instead of constructing confidence intervals for $f^{\prime}(\theta)$, SiZer seeks confidence intervals for the scale--space version $f^{\prime}(\theta;\nu)\equiv \mathbb{E}(\hat{f}^{\prime}(\theta;\nu))$. As usual, confidence limits are of the form
\begin{equation}
\hat{f}^{\prime}(\theta;\nu)\pm q \cdot \widehat{\mbox{sd}}(\hat{f}^{\prime}(\theta;\nu)),
\label{interval}
\end{equation}
where $q$ is an appropiate quantile and $\widehat{\mbox{sd}}$ is the estimated standard deviation (details on its computation are given below).

So, at each pair $(\theta,\nu)$, if 0 lies in the corresponding interval, the slope of the smoothed curve is not significant, whereas positive and negative intervals will indicate increasing and decreasing trends. Behaviour at $\theta$ and $\nu$ will be presented via CircSiZer color map, as discussed in Section \ref{reading_sizer}.

In the linear case, Chaudhuri and Marron (1999) suggested several methods for the approximation of the quantile $q$, including pointwise and simultaneous Gaussian quantiles and also bootstrap quantiles. In our setting, accurate intervals without imposing Gaussian assumptions in (\ref{interval}) can be obtained by bootstrap (see Efron and Tibshirani, 1993). A possible way to get such intervals, namely the ``bootstrap--$t$'' approach, is detailed below. Given a significance level $\alpha$ and for a fixed value of $\nu>0$ and with $\theta$ varying in the interval $[0,2\pi)$, the following algorithm is considered:
\begin{itemize}
\item[Step 1.] Generate $B$ bootstrap samples, i.e., random samples drawn with replacement from the data.
\item[Step 2.] For each bootstrap sample, compute
\[
Z^{*}_i \equiv Z^{*}_i(\theta;\nu)=\frac{\hat{f}^{\prime}(\theta;\nu)^{*} - \hat{f}^{\prime}(\theta;\nu)}{\widehat{\mbox{sd}}(\hat{f}^{\prime}(\theta;\nu)^{*})},\ i=1,\ldots,B.
\]
where $\hat{f}^{\prime}(\theta;\nu)^{*}$ is the value of $\hat{f}^{\prime}(\theta;\nu)$ for the bootstrap sample and $\widehat{\mbox{sd}}(\hat{f}^{\prime}(\theta;\nu)^{*})$ is 
an estimator of the standard deviation of $\hat{f}^{\prime}(\theta;\nu)^{*}$ (see Section \ref{standard_deviation} for its calculation).
\item[Step 3.] Based on $Z^*_1,\ldots, Z^*_B$ compute the $\alpha$ and $(1-\alpha)$ sample quantiles, $\hat{t}^{(\alpha)}$ and $\hat{t}^{(1-\alpha)}$, respectively. 
\item[Step 4.] The ``bootstrap--$t$'' confidence interval is given by
\[
\left(\hat{f}^{\prime}(\theta;\nu)-\hat{t}^{(1-\alpha)}\cdot \widehat{\mbox{sd}}(\hat{f}^{\prime}(\theta;\nu)), \hat{f}^{\prime}(\theta;\nu)-\hat{t}^{(\alpha)}\cdot \widehat{\mbox{sd}}(\hat{f}^{\prime}(\theta;\nu))\right),
\]
where $\widehat{\mbox{sd}}(\hat{f}^{\prime}(\theta;\nu))$ is an estimator of the standard deviation of $\hat{f}^{\prime}(\theta;\nu)$ (see Section \ref{standard_deviation} for its calculation).
\end{itemize}

\subsection{Estimation of the standard deviation}
\label{standard_deviation}

For the computation of confidence intervals, it is necessary to derive an expression for $\widehat{\mbox{sd}}(\hat{f}^{\prime}(\theta;\nu))$ (and also for its bootstrap version $\widehat{\mbox{sd}}(\hat{f}^{\prime}(\theta;\nu)^*)$, involved in the standarization procedure in Step 2 of the previous algorithm). The main idea behind the calculation, in the context of density estimation, is that the derivative estimator $\hat{f}^{\prime}(\theta;\nu)$ is a weighted average of the derivative of the kernel function at different locations. Specifically, for the problem of density estimation and following Chaudhuri and Marron (1999), our proposal is to estimate the variance of $\hat{f}^{\prime}(\theta;\nu)$ by

\[
\begin{array}{lll}
\widehat{\mbox{var}}\left(\hat{f}^{\prime}(\theta;\nu)\right) & = & \widehat{\mbox{var}}\left(n^{-1}\sum_{i=1}^{n}K_{\nu}^{\prime}(\theta-\Theta_i)\right) \\
 & & \\
& = & n^{-1}s^{2}\left(K_{\nu}^{\prime}(\theta-\Theta_1),\ldots,K_{\nu}^{\prime}(\theta-\Theta_n)\right), \ \ 0\leq\theta<2\pi,
\end{array}
\]
where $s^2$ is the usual sample variance of $n$ data, which in this context is formed by the derivative of the kernel centered at each sample value $\Theta_i$, with $i=1,\ldots,n$.

In the regression setting, the derivative estimator is given by $\hat{f}^{\prime}(\theta;\nu)=\hat{b}$, see (\ref{regression}). It can be shown (see, for instance, Wasserman (2006), p.77) that $\hat{f}^{\prime}(\theta;\nu)$ can be written as
$$
\hat{f}^{\prime}(\theta;\nu)=\frac{1}{n}\sum_{i=1}^n W_{\nu}(\theta,\Theta_i)Y_i,
$$
for some certain weights $W_{\nu}(\theta,\Theta_i)$ which can be easily computed from the kernel $K_{\nu}$. So, the variance of $\hat{f}^{\prime}(\theta;\nu)$ is given by:

\[
\begin{array}{lll}
\mbox{var}\left(\hat{f}^{\prime}(\theta;\nu)\right) & = &\mbox{var}\left(n^{-1}\sum_{i=1}^{n} W_{\nu}(\theta,\Theta_i)Y_i | \Theta_1,\ldots,\Theta_n\right) \\
& & \\
& = & \sum_{i=1}^{n} \sigma^{2}(Y_i|\Theta_i)(W_{\nu}(\theta,\Theta_i))^2.\\
\end{array}
\]

A crucial problem is how to estimate the conditional variance $\sigma^{2}(Y_i|\Theta_i)$. When the covariate is linear, Chaudhuri and Marron (1999) estimate this quantity by smoothing the residuals using the same ``linear'' bandwidth as the one used to calculate the estimator.  When the covariate is circular, the bandwidth used to compute the estimator is devised for a circular framework, whereas the residuals are linear. Hence, it does not seem reasonable to smooth the residuals using the same bandwidth $\nu$, which comes from the circular setting. In order to avoid the calculation of a new bandwidth for smoothing the residuals, the standard deviation will be approximated by bootstrap. For a given $\nu>0$ and  with $\theta$ varying in $[0,2\pi)$, the standard deviation of $\hat{f}^{\prime}(\theta;\nu)$ is estimated following the next steps:
\begin{enumerate}
\item Generate $B$ bootstrap samples, each one consisting of $n$ data values drawn with replacement from the observed sample $\left\{(\Theta_i,Y_i);\, i=1,\ldots,n\right\}$.
\item For each bootstrap sample, calculate $\hat{f}^{\prime*b}(\theta;\nu)$, with $b=1,\ldots,B$.
\item Estimate the standard deviation by the sample standard deviation of the $B$ replicates:
\[
\begin{array}{lll}
\widehat{\mbox{sd}}\left(\hat{f}^{\prime}(\theta;\nu)\right) & = & \left[s^2\left(\hat{f}^{\prime*1}(\theta,\nu),\ldots,\hat{f}^{\prime*B}(\theta,\nu)\right)\right]^{1/2} .\\
\end{array}
\]
\end{enumerate}

\subsection{Reading CircSiZer}
\label{reading_sizer}

As noted above, with CircSiZer, significance features in the data will be seeked via the construction of confidence intervals for the scale--space version of the smoothed derivative curve. Although the procedure for obtaining these intervals must be carefully adapted for circular data involved in density estimation and introduced as covariates in regression estimation with linear response, as shown along this section, the interpretation of the output through CircSiZer map is fairly simple.

Recall that, for a given pair $(\theta,\nu)$, the curve at a smoothing level $\nu$ is significantly increasing (decreasing) if the confidence interval is above (below) 0 and if the confidence interval contains 0, the curve at the smoothing level $\nu$ and at the point $\theta$ does not have a statistically significant slope. This information can be displayed in a circular color map in such a way that, at a given $\nu$, the performance of the estimated curve is represented by a color ring with radius proportional to $\nu$. Differents colors will allow to indentify peaks and valleys.

Blue (black, for black and white versions) color indicates locations where $f(\theta,\nu)$ is significantly increasing; red (dark gray) color shows where it is significantly decreasing and purple (gray) indicates where it is not significantly different from zero. There is also a fourth color, gray (light gray), corresponding to those regions where there is not enough data to make statements about significance. Thus, at a given bandwidth, a significant peak can be identified when a region of significant positive gradient is followed by a region of significant negative gradient (i.e. blue--red pattern), and a significant trough by the reverse (red--blue pattern), taking clockwise as the positive sense of rotation. In Section \ref{applications}, some examples of CircSiZer map with simulated and real data are shown.

To determine the gray areas (not enough data), for each $(\theta,\nu)$ the estimated effective sample size (ESS) is calculated as
\[
\mathrm{ESS}(\theta,\nu)=\frac{\sum_{i=1}^{n}K_{\nu}(\theta-\Theta_i)}{K_{\nu}(0)}.
\]
Following Chaudhuri and Marron (1999), regions where $\mathrm{ESS}(\theta,\nu)<5$ are shaded gray.

\section{Examples and real data analysis}
\label{applications}

In this section, the performance of CircSiZer for density and regression is illustrated with some simulated examples and a real dataset. The corresponding CircSiZer maps were obtained from self--programed code in R (see R Development Core Team, 2012), which is available as supplementary material.

\begin{figure}[h!]
\begin{center}
\includegraphics[width=5cm]{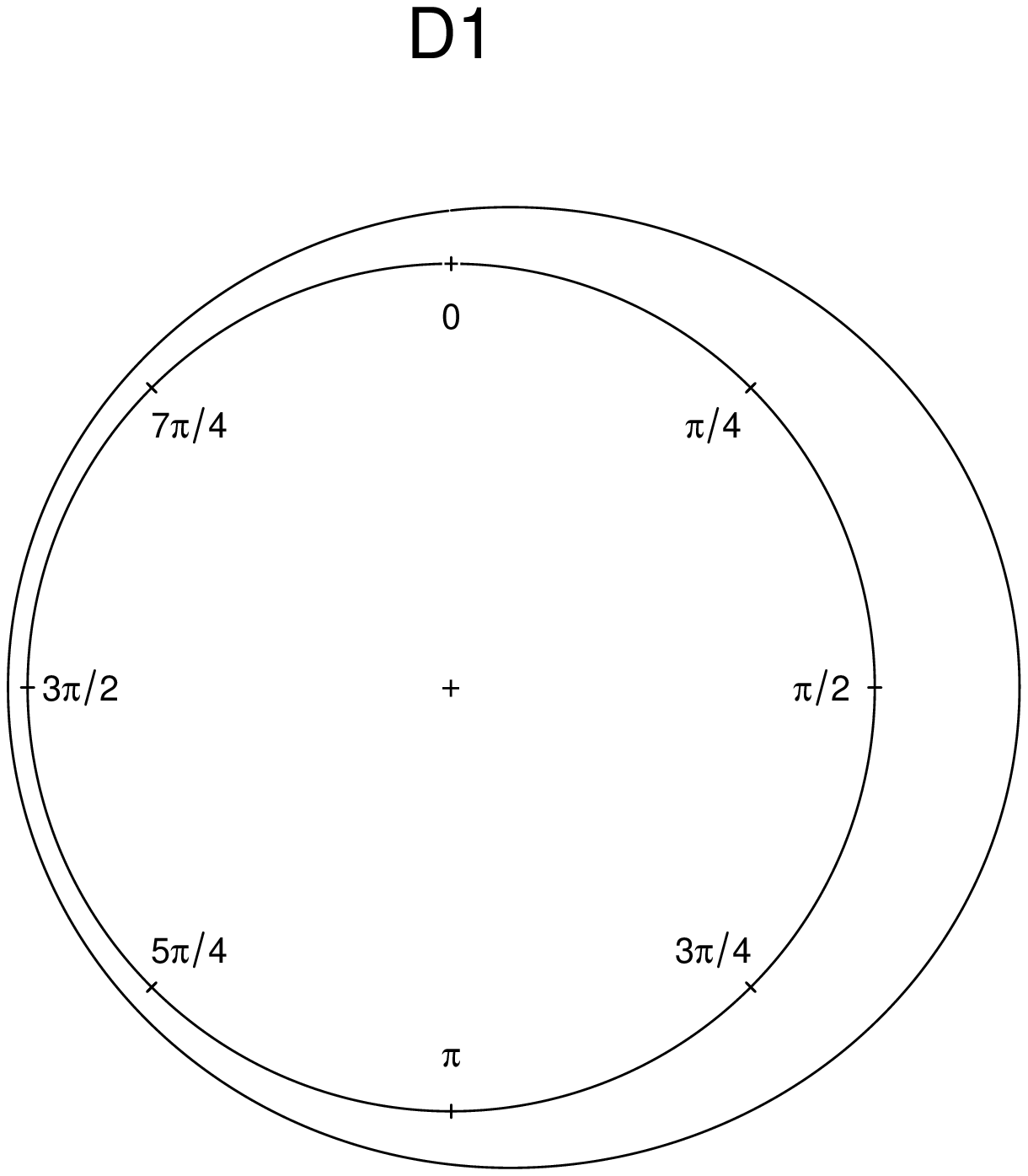}
\includegraphics[width=5cm]{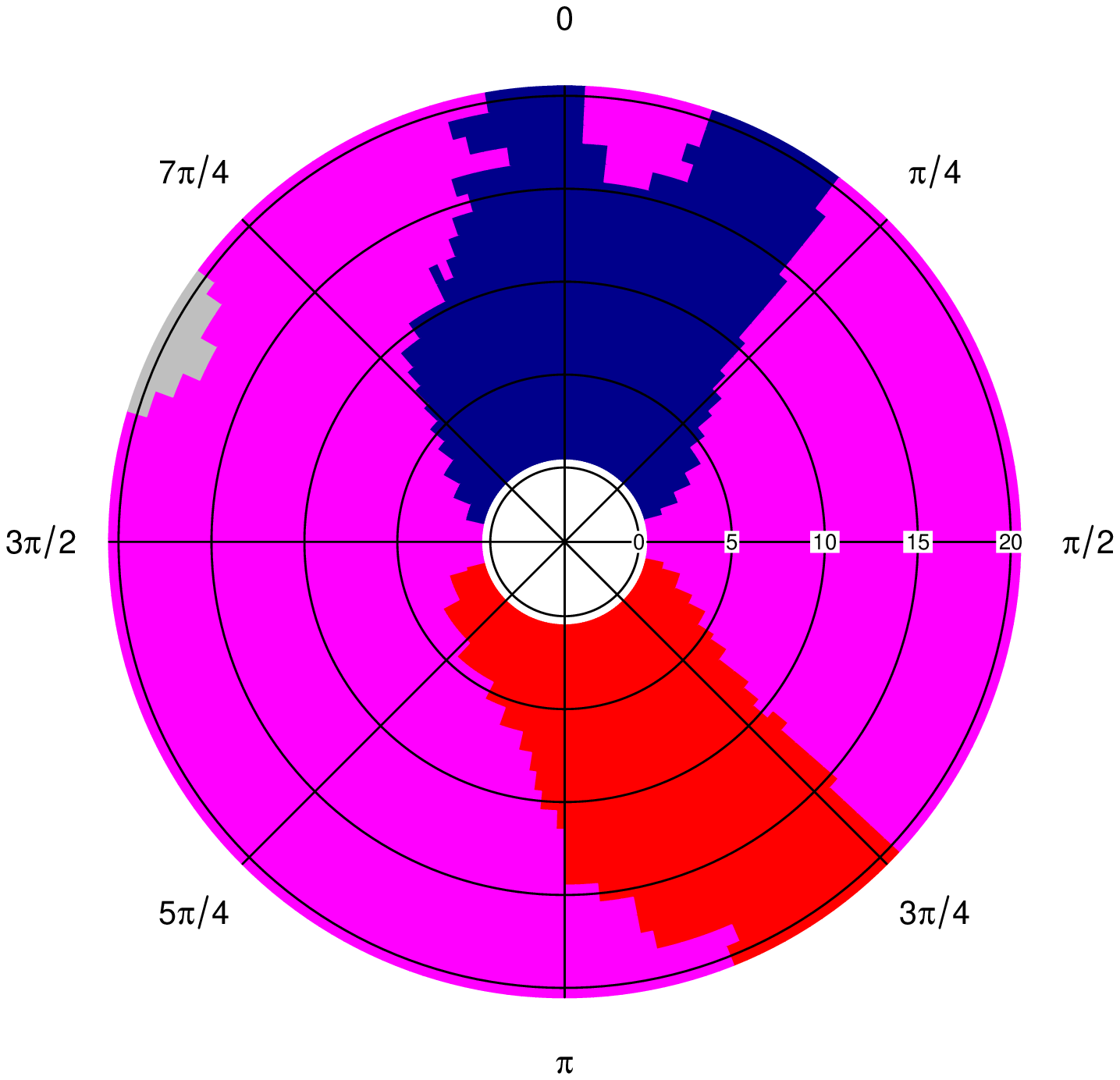}\\ 
\includegraphics[width=5cm]{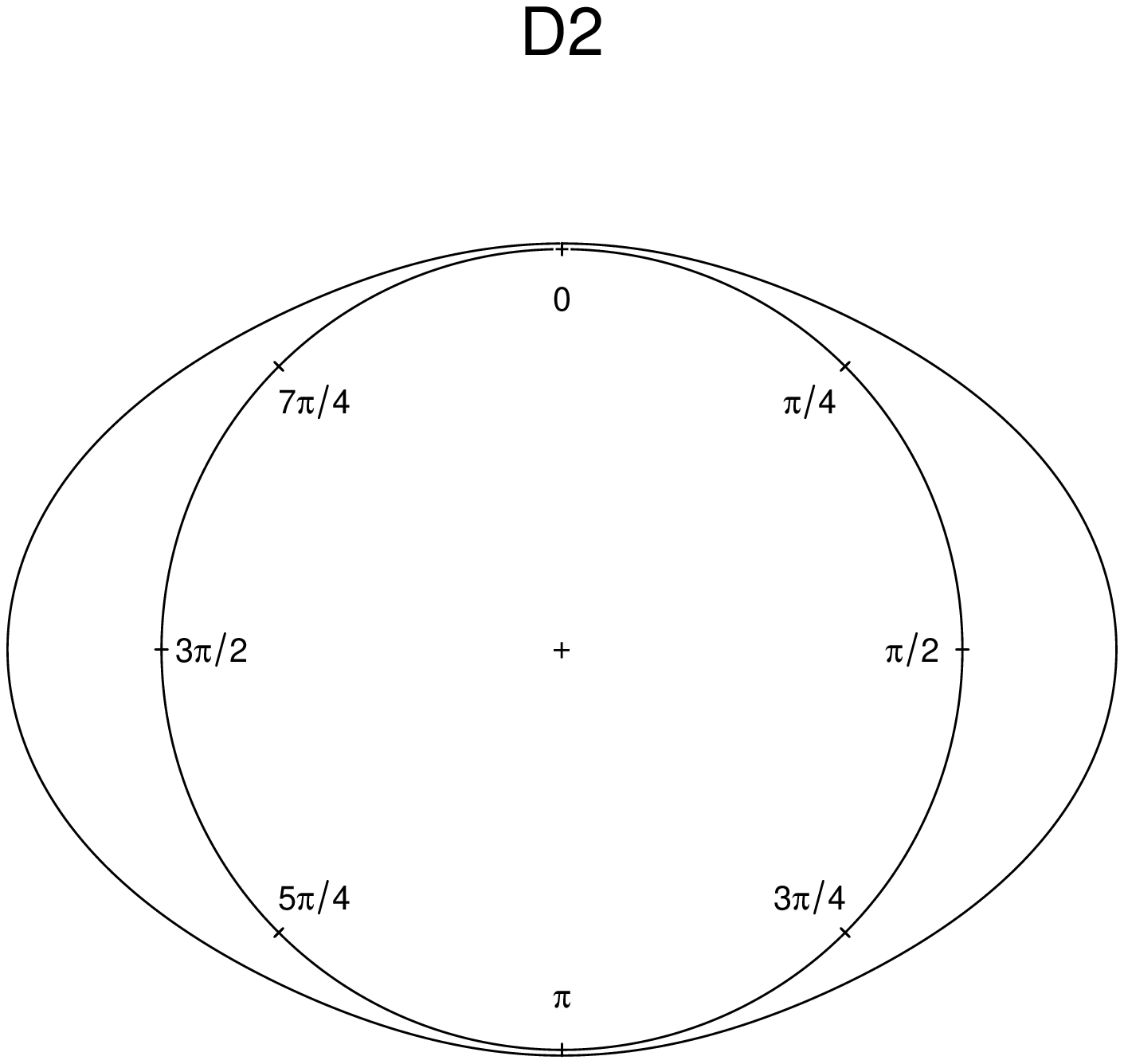}
\includegraphics[width=5cm]{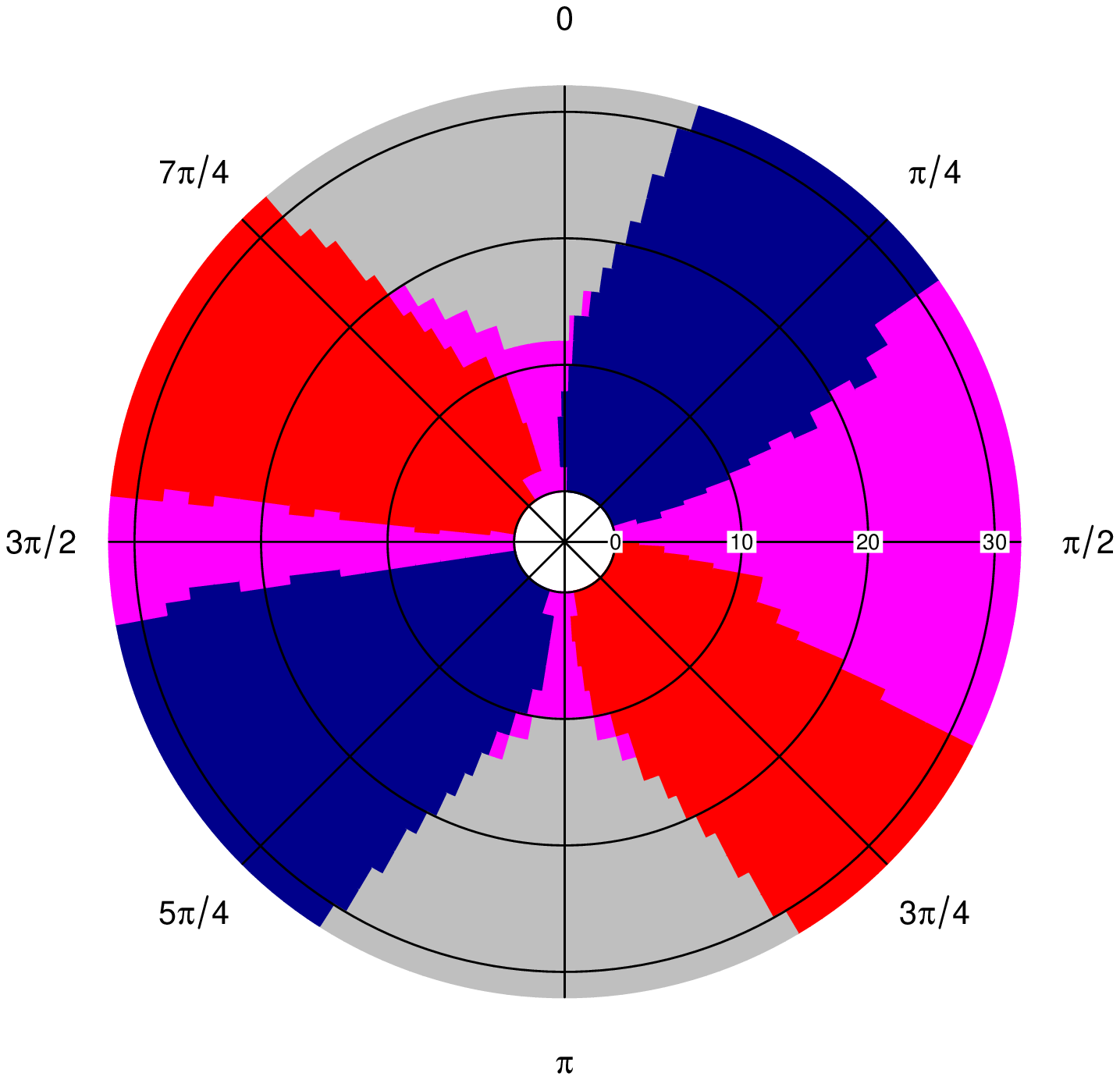}\\ 
\includegraphics[width=5cm]{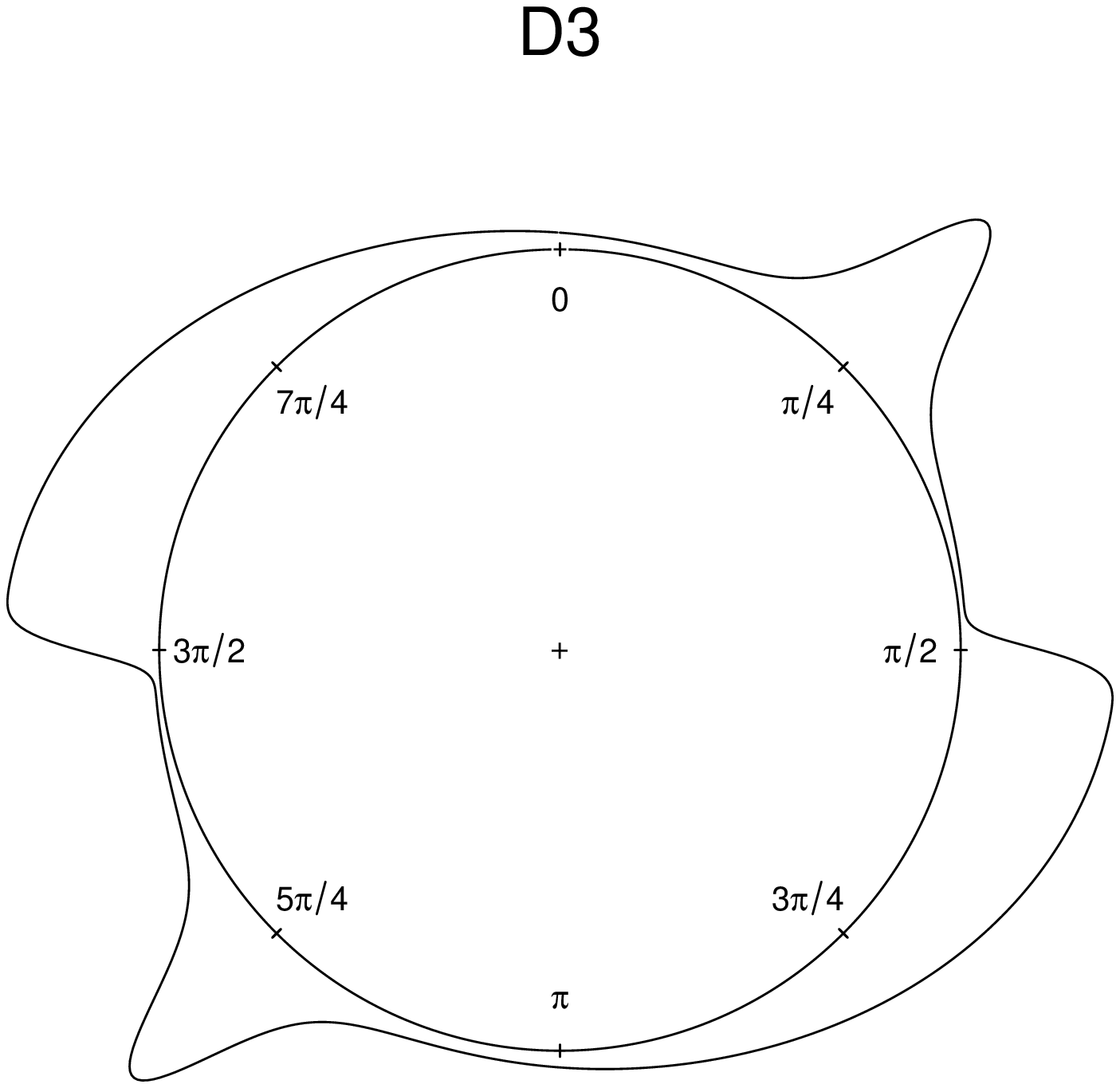}
\includegraphics[width=5cm]{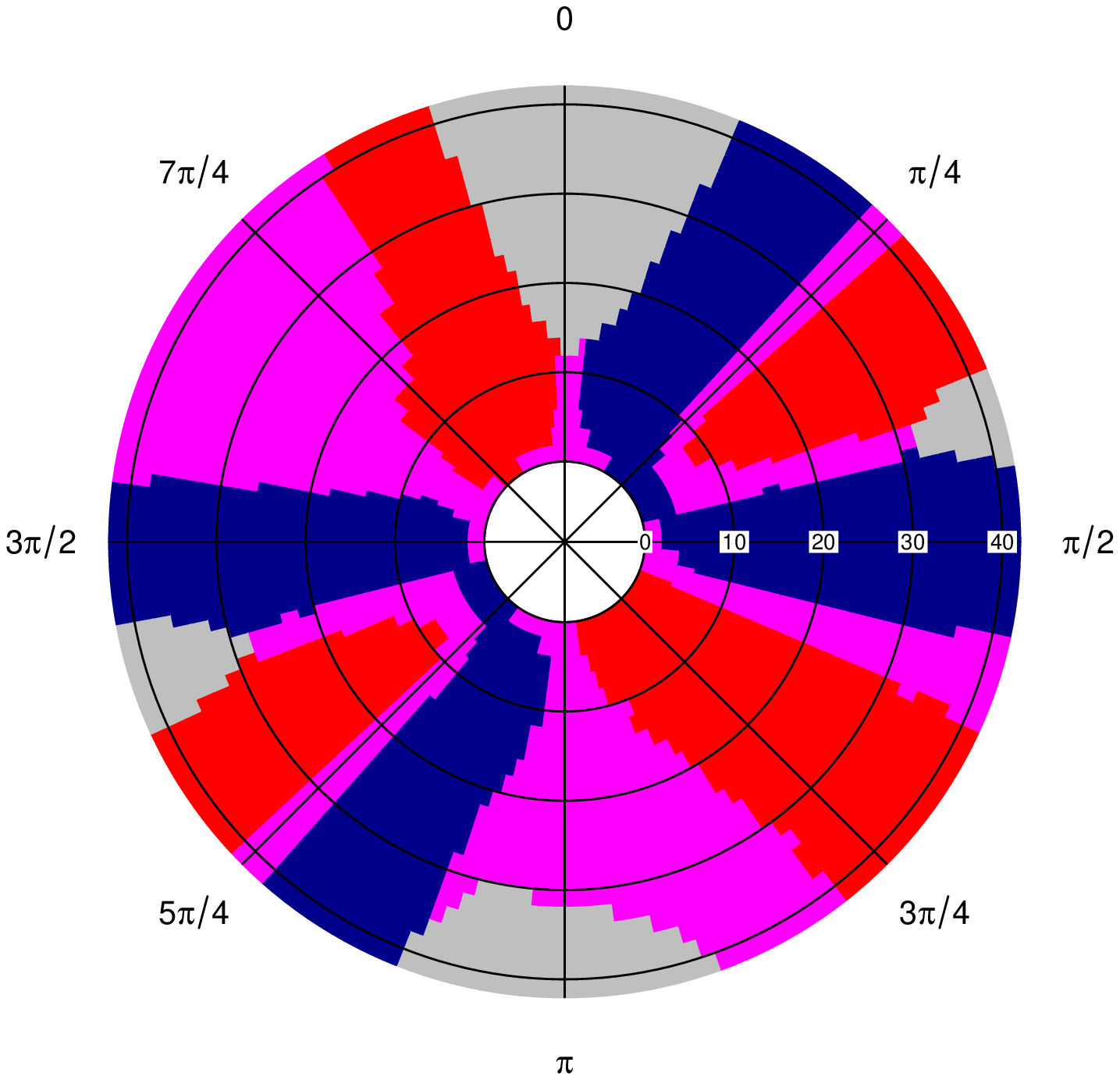}\\ 
\caption{CircSiZer maps (right column) for kernel density estimates based on simulated data from densities, D1, D2 and D3 (left column). Sample size $n=200$. For reading CircSiZer, take clockwise sense of rotation. Values of $\nu$ are indicated along the radius.}
\label{sizer}
\end{center}
\end{figure}

\begin{figure}[h!]
\begin{center}
\includegraphics[width=5cm]{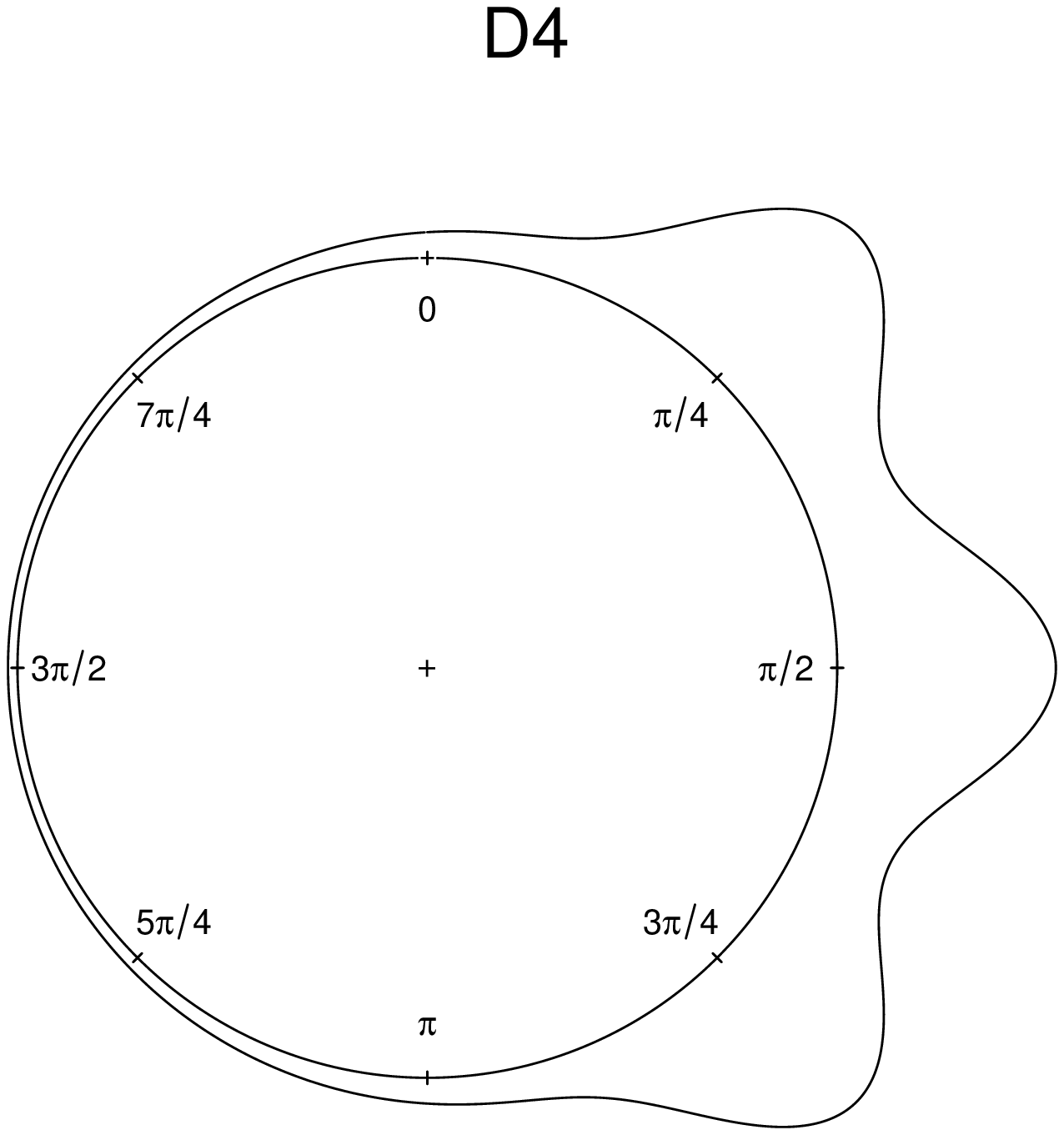}\\
\includegraphics[width=5cm]{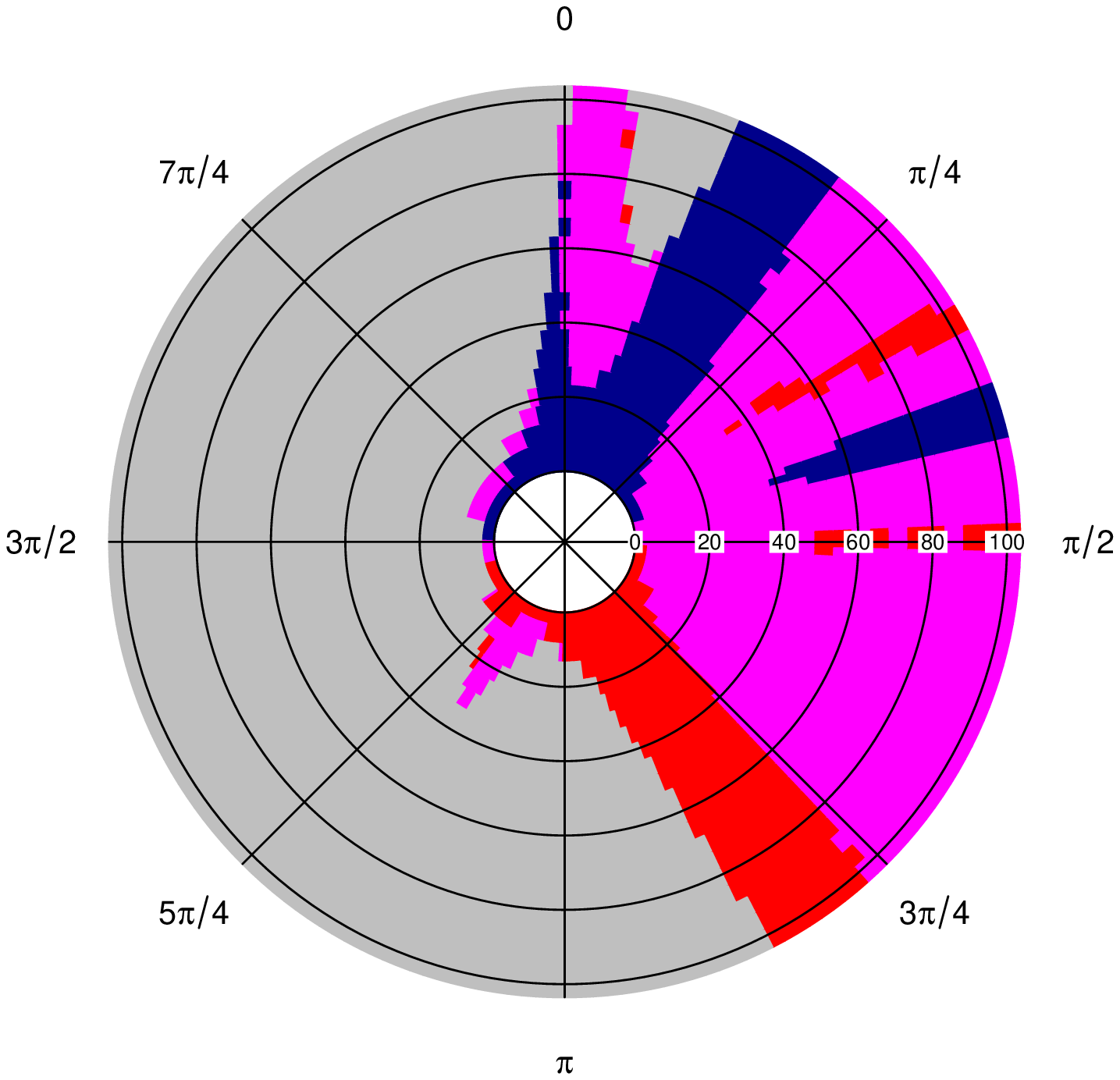} 
\includegraphics[width=5cm]{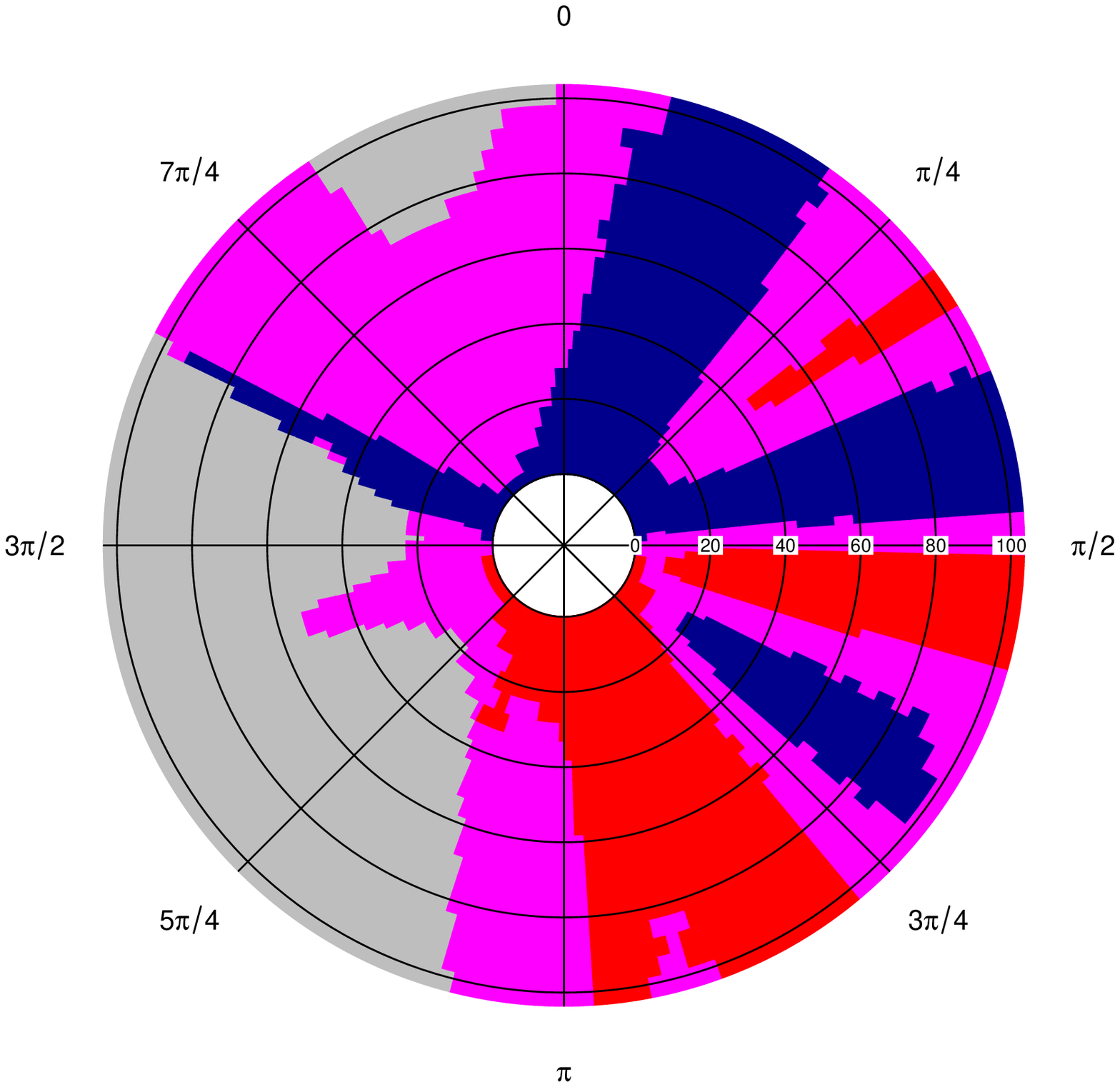}
\caption{CircSiZer maps for kernel density estimates based on simulated data with sample size $n=200$ (bottom--left) and $n=500$ (bottom--right) from density D4 (top row). For reading CircSiZer, take clockwise sense of rotation. Values of $\nu$ are indicated along the radius.}
\label{sizer2}
\end{center}
\end{figure}

For density estimation, the performance of CircSiZer has been studied in some simulated scenarios, which highlight how CircSiZer displays the information available in the data. Several circular distributions have been considered: von Mises (D1), mixture of two and four von Mises (D2, D4) and mixture of two wrapped Cauchy and two wrapped skew--normal distributions (D3). See Figure \ref{sizer} (left column) and Figure \ref{sizer2} (top row) for density plots and Oliveira et al. (2012a) for specific formulae. Throughout this section, statistical significance is assessed with a significance level $\alpha=0.05$.

Figure \ref{sizer} (rigth column) presents CircSiZer maps for densities, D1, D2 and D3, with random samples of size $n=200$. The CircSiZer maps can be easily obtained by calling the function {\tt circsizer.densi-}\\ {\tt ty(x,NU,ngrid,alpha,B,type)}. The arguments in this function are {\tt x}, the angle data sample; {\tt NU}, a grid of positive smoothing parameters and {\tt ngrid}, an integer indicating the number of equally spaced angles between 0 and $2\pi$ where the estimator is evaluated (default to {\tt ngrid=}$250$). A significance level {\tt alpha} can be also fixed (default to {\tt alpha=}$0.05$), as well as the number of bootstrap samples {\tt B} to estimate the standard deviation of $\hat{f}^{\prime}(\theta;\nu)$ (default to {\tt B=}$500$). Finally, {\tt type} is a number indicating the labels that appear in the plot: 1 (directions), 2 (hours), 3 (angles in radians) or 4 (angles in degrees). Default is {\tt type=3}. It is clear that the CircSiZer maps show the significance of the unimodal, bimodal and cuatrimodal structure for each density, respectively. Taking clockwise as the positive sense of rotation, Figure \ref{sizer} (top--right) displays a blue area followed by a red area for a wide range of bandwidths, indicating a significant increase then decrease, i.e., unimodality. In Figure \ref{sizer} (center--right), the bimodal structure is clearly brought out by the CircSiZer map, as the two peaks and the trough can be identified by the clockwise blue--red--blue--red pattern on the map that occurs for a range of bandwidths between $\nu=1$ and $\nu=31$. In Figure \ref{sizer} (bottom--right), it can be seen that only two modes are identified for values of the smoothing parameter smaller than $\nu=10$ but, for larger values of this parameter the cuatrimodal structure is obvious.

The effect of increasing the sample size $n$ in model D4 can be seen in Figure \ref{sizer2}. For $n=200$, CircSiZer map detects only two signficant modes (see Figure \ref{sizer2}, bottom--left). However, the underlying three modes are significant for $n=500$ (Figure \ref{sizer2}, bottom--right).

\begin{figure}[h!]
\begin{center}
\includegraphics[width=5cm]{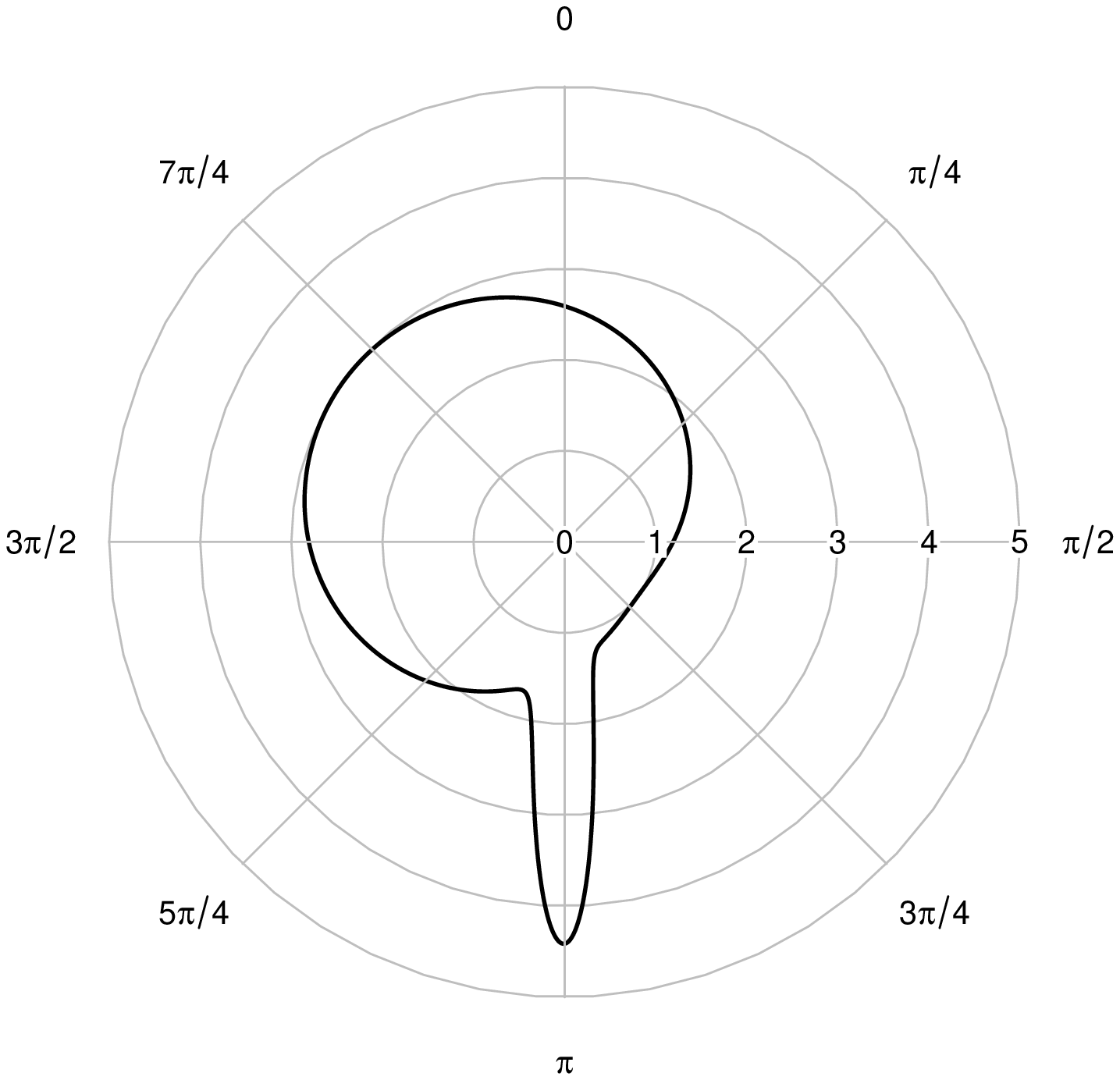}
\includegraphics[width=5cm]{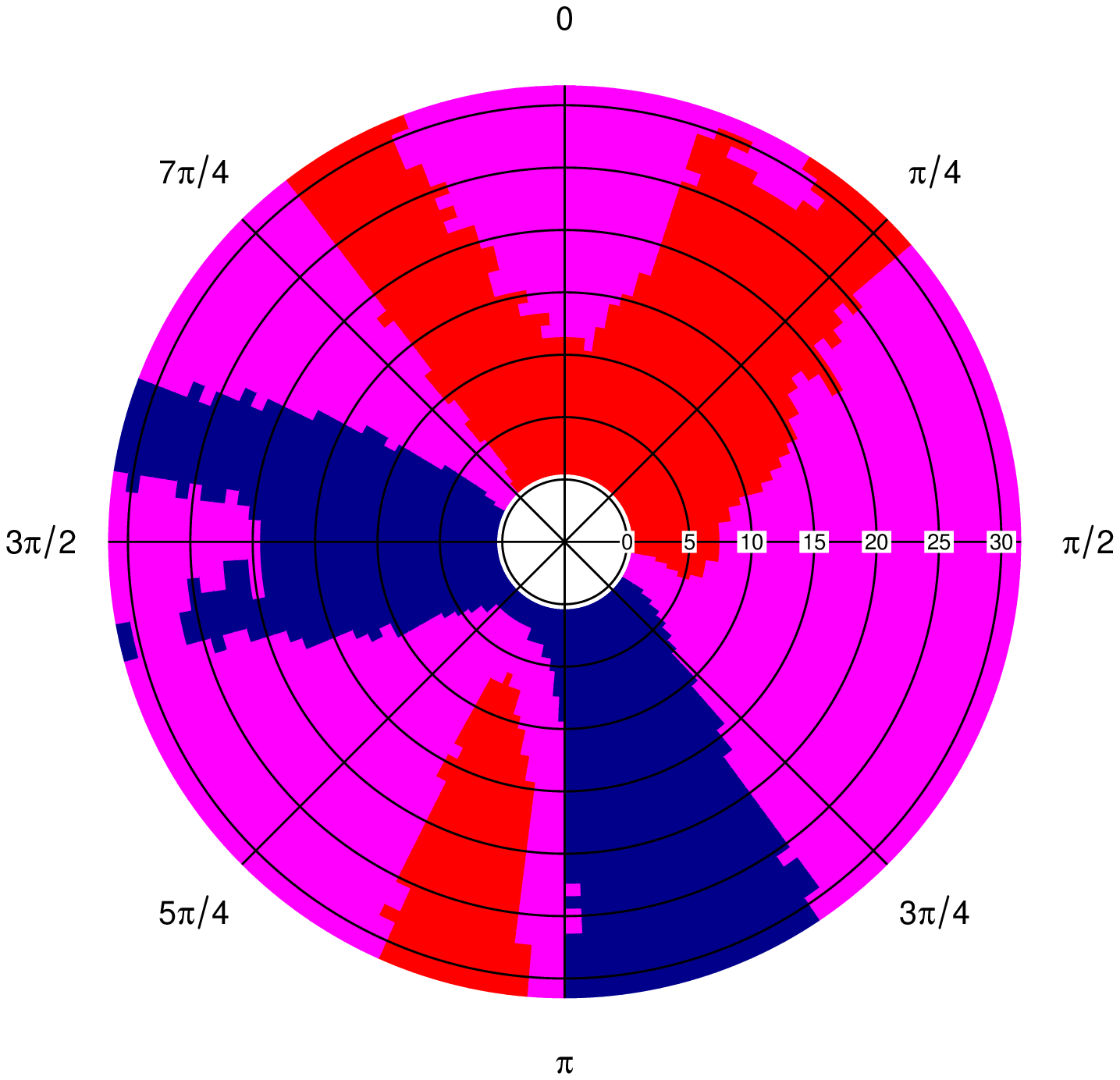}\\
\caption{CircSiZer map for kernel regression estimate based on simulated data with sample size $n=200$ from model (\ref{regression_model}). The regression function is plotted in the left panel. For reading CircSiZer, take clockwise sense of rotation. Values of $\nu$ are indicated along the radius.}
\label{sizer3}
\end{center}
\end{figure}

For illustrating the performance of CircSiZer in estimating a regression model such as (\ref{regression_model}), the regression function displayed in Figure \ref{sizer3} (left panel) has been considered. This is the same model already analyzed by Di Marzio et al. (2009) and the fourth model in the illustration of kernel estimators presented by Oliveira et al. (2012b). A sample of 200 observations from model (\ref{regression_model}), with normally distributed errors with variance $\sigma^2=0.5$, has been generated in order to produce the CircSiZer map. 

In the regression setting, the CircSiZer map can be obtained by calling the function {\tt circsizer.re-}\\ {\tt gression(x,y,NU,ngrid,alpha,B,B2,type)}. The first arguments for this function are {\tt x}, the circular covariate values and {\tt y}, the linear response vector. The arguments {\tt NU}, {\tt ngrid}, {\tt alpha}, {\tt B} and {\tt type} are the same as for the density case, with the same default values for {\tt alpha}, {\tt B} and {\tt type}. Default for {\tt ngrid=}$150$. For the regression CircSiZer, a further argument {\tt B2} is required. {\tt B2} is the number of bootstrap samples used to compute the denominator in Step 2 of the algorithm (default is {\tt B2}=$250$). From Figure \ref{sizer3} (left panel), it is clear that the regression model to estimate may present some challenges given the highly peaked mode centered in $\pi$ and another less concentrated mode about $7\pi/4$. Nevertheless, as it can be seen in the CircSiZer map presented in Figure \ref{sizer3} (right panel), the two modes are identified as significant along the range of bandwidths.

\subsection{Exploring wind patterns using CircSiZer}
\label{real_data}
The practical usefulness of the proposed CircSiZer map is illustrated by the analysis of a real dataset concerning wind direction and speed in the atlantic coast of Galicia (NW--Spain). Meteorological and oceanographic variables related to wind and currents behaviour are collected by a standard buoy (model SeaWatch). With a diameter of 1.8m and a height of 6.5m, the buoy is anchored at the location specified in Figure \ref{coast_map}, far away from the coastline so that the measurements are not influenced by local effects. Wind measurements regarding direction and speed are recorded every ten minutes, and hourly averaged, at a height of 3m above sea level. Data can be freely downloaded from the Spanish Portuary Authority (Puertos del Estado, http://www.puertos.es).

The dataset consists of hourly observations of wind direction (in degrees) and wind speed (in m/s) in winter season (from November to February), from 2003 until 2012. For the circular representation, as in previous plots, wind direction is marked over the circumference clockwise, starting from N. In order to avoid the dependence present between consecutive measurements in the time series, the autocorrelation functions were studied. Observations taken with a lag period of 95 hours can be considered as uncorrelated, providing a final dataset with about 200 values. With this lag period, all the day hours are represented in the sample.

\begin{figure}[h!]
\begin{center}
\includegraphics[width=5cm]{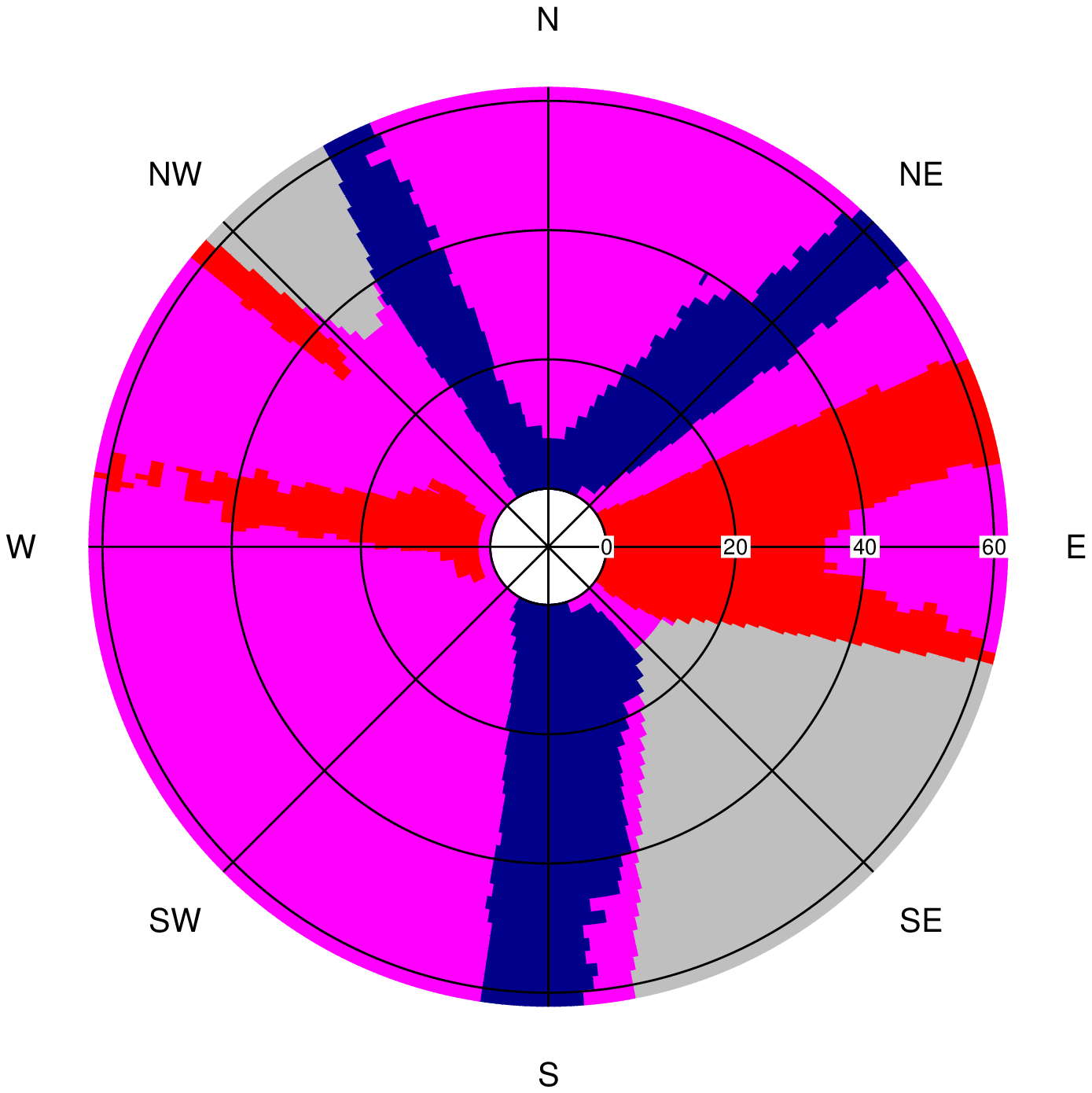}
\includegraphics[width=5cm]{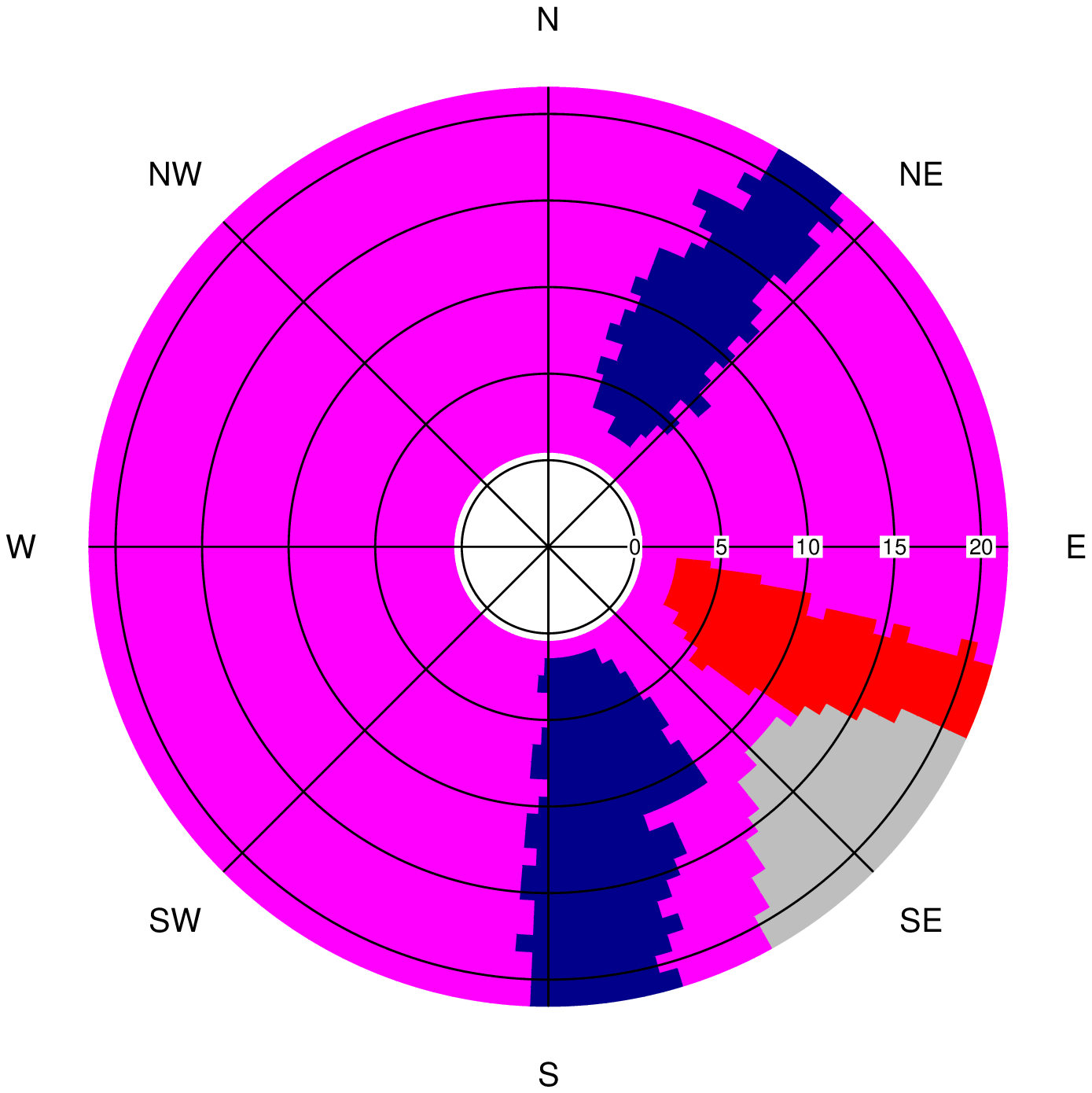}
\caption{CircSiZer map for kernel density estimator (left) for wind direction and CircSiZer map for circular--linear regression (right) for wind speed (m/s) with respect wind direction. For reading CircSiZer, take clockwise sense of rotation. Values of $\nu$ are indicated along the radius.}
\label{real_data_fig}
\end{center}
\end{figure}

Figure \ref{real_data_fig} shows the CircSiZer maps for wind directions (left plot) and CircSiZer for regression (right plot), applied for exploring the relation between wind speed as a response and wind direction as a covariate. In Figure \ref{real_data_fig} (left plot), the two significant modes that can be distinguished for a wide range of bandwidths indicate that winds in winter period come mostly from NE and SW. Winds from SE are not frequent at all, being this fact reflected by the absence of data in the SE sector (gray shaded area). In addition, it can be also seen that wind speed increases when wind direction comes from NE and S (Figure \ref{real_data_fig}, right plot).

\section*{Final comments}
\label{discussion}

An extension of SiZer to circular data, the CircSiZer, both for density and regression, has been proposed. The performance of CircSiZer has been checked by some simulated examples and it has also been applied to analyze wind patterns in Galician coast during winter season. In order to effectively produce the CircSiZer map, the assessment of the variability in the derivatives of the circular kernel estimators, both for density and regression, is approached through the computation of standard deviations and appropriate quantiles by bootstrap methods. Despite the technical details behind the CircSiZer derivation, possibly overwhelming for a practitioner, the graphical output appearance allows for an easy and useful interpretation.

As mentioned in the introduction, the SiZer technique has been adapted to other settings. Although most of the previous works, and also the proposal presented in this paper, consider smoothers based on kernels, the technique could be adapted for other type of smoothers such as splines (Marron and Zhang, 2005). The same extension could be possible for circular data, although suitable modifications should be done in order to account for the periodic nature of the data.

It should be noted that circular data are just a particular case of spherical data (data on the $q$--dimensional sphere). In principle, the methodology presented in Section 3 could be extended to higher dimensions. Nevertheless, the lack of a simple visualization device will certainly hampered the practical purpose of CircSiZer for general dimension.

Finally, self--programmed code has been implemented for applying the proposed methods in practice. This code, developed in R (R Development Core Team, 2012), is available as supplementary material.

\section*{Acknowledgements}
This work has been supported by Project MTM2008--03010 from the Spanish Ministry of Science and Innovation, and by the IAP network StUDyS (Developing crucial Statistical methods for Understanding major complex Dynamic Systems in natural, biomedical and social sciences), from Belgian Science Policy. We also acknowledge the advice of Jos\'e A. Crujeiras, an experienced skipper working in the Galician coast.



\end{document}